\documentclass[fleqn,usenatbib]{mnras}

\usepackage[T1]{fontenc}

\DeclareRobustCommand{\VAN}[3]{#2}
\let\VANthebibliography\thebibliography
\def\thebibliography{\DeclareRobustCommand{\VAN}[3]{##3}\VANthebibliography}

\usepackage{graphicx}	
\usepackage{amsmath}	
\usepackage{amssymb}	
\usepackage{newtxtext,newtxmath}
\usepackage{xcolor,colortbl}
\definecolor{Gray}{gray}{0.9}
\definecolor{silver}{rgb}{0.75, 0.75, 0.75}
\newcommand{\RomanNumeralCaps}[1]
    {\MakeUppercase{\romannumeral #1}}

\title[Study of extragalactic BH-XRBs]{Accretion disc dynamics in extragalactic black hole X-ray binaries: A comprehensive study of M33 X-7, NGC 300 X-1 and IC 10 X-1}

\author[Bhuvana \& Nandi 2024]{
Bhuvana G. R.,$^{1,2}$\thanks{E-mail: bhuvanahebbar@gmail.com /\ bhuvana.g.r.k4@f.gifu-u.ac.jp}
and Anuj Nandi$^{3}$
\\
$^{1}$Department of Physics, Dayananda Sagar University, Bengaluru, 560068\\
$^{2}$Institute for Advanced Study, Gifu University, 1-1 Yanagido, Gifu 501-1193, Japan \\
$^{3}$Space Astronomy Group, ISITE Campus, U R Rao Satellite Centre, Bengaluru, 560037\\
}

\date{Accepted XXX. Received YYY; in original form ZZZ}

\pubyear{2015}

\begin{document}
\label{firstpage}
\pagerange{\pageref{firstpage}--\pageref{lastpage}}
\maketitle

\begin{abstract}
Extragalactic Black Hole X-ray Binaries (BH-XRBs) are the most intriguing X-ray sources as some of them are `home' to the most massive stellar-mass BHs ever found. In this work, we conduct a comprehensive study of three massive, eclipsing extragalactic BH-XRBs i.e., M33X-7, NGC300X-1, and IC10X-1 and using entire X-ray observations available from \textit{XMM-Newton} and \textit{NuSTAR} till date. Preliminary analysis using \textit{diskbb} and \textit{powerlaw} models shows that the sources have steep spectra and sub-Eddington luminosities (L$<$0.69 L$_{Edd}$), with major flux contribution from non-thermal component, resembling the relatively uncharted Steep Powerlaw State (SPL). To understand the accretion disc properties in this state, we explore alternate modelling scenario that reveals the presence of a `hot' ($kT_{in}=1-2$ keV) slim-disc (\textit{diskpbb}) with radial temperature profile $T(r)\propto r^{-p}$ ($p=0.5-0.66$), along with a cooler ($kT_{in}=0.1-0.2$ keV) standard thermal disc (\textit{diskbb}). We carry out the continuum-fitting method using relativistic slim-disc model (\textit{slimbh}) and estimate the mass range of M33 X-7, NGC300X-1 and IC10X-1 is to be 9$-$15 M$_{\odot}$,  9$-$28 M$_{\odot}$ and 10$-$30 M$_{\odot}$, respectively. Further, eclipse periods are determined by modelling the lightcurve, using which we estimate the size of the eclipsing bodies. Modelling of the eclipse spectra revealed the complete obscuration of soft spectral component during eclipse, implying the emission of hard component from an extended accretion region. Based on our findings, we provide an inference on geometry of accretion disc in these wind-fed systems and compare their properties with the other two extragalactic BH-XRBs.

\end{abstract}




\section{Introduction}
\label{sec1}
The X-ray binary (XRB) systems consist of a  stellar mass Black Hole (BH) along with a  non-degenerate companion star. A significant number of stellar mass XRBs are found in our Galaxy, which includes transient as well as persistent sources. Transients are the one that remain in quiescence majority of the time and undergo occasional outbursts, while persistent sources maintain consistent emission in X-rays \citep{2016A&A...587A..61C,2016ApJS..222...15T,2019MNRAS.487..928S}. BH-XRBs detected outside our Galaxy include Ultra-Luminous X-ray sources (ULXs) besides stellar mass XRBs analogous to the Galactic XRBs. The ULXs are extremely luminous (L $>10^{39}$ erg s$^{-1}$) X-ray binaries that host either a stellar mass black hole/neutron star accreting at super-Eddington rate or an intermediate mass black hole with sub-Eddington accretion (\citealt{2011NewAR..55..166F,2015MNRAS.446.3926A,2017ARA&A..55..303K,2021AstBu..76....6F,2023MNRAS.526.2086M}). Stellar mass ULXs with super-Eddington accretion are discovered to be present in large number outside our Galaxy \citep{2022MNRAS.509.1587W}. Meanwhile, only a handful of sub-Eddington stellar-mass extragalactic BHs are found till date, which includes M33 X-7 located in M33 \citep{1988ApJ...325..531T}, NGC 300 X-1 in NGC 300 \citep{2001A&A...373..473R}, IC 10 X-1 in IC 10 \citep{1997MNRAS.291..709B}, LMC X-1 and LMC X-3 in Large Magellanic Cloud (LMC) \citep{1979ApJ...233..514J} galaxies. All of these sources accrete via stellar wind from its massive companion stars except LMC X-3, which accretes via Roche Lobe Overflow (RLOF) \citep{1983ApJ...272..118C}. The X-ray emission from these sources are persistently bright with luminosity in the range of $10^{36}-10^{39}$ erg s$^{-1}$. These characteristics set them apart as unique sources to investigate accretion dynamics around black hole sources. 

Spectral analysis of BH-XRBs serves as a vital tool to investigate accretion geometry as it allows us to understand the properties of various spectral components originating from different regions around the BH. An X-ray spectrum of a typical BH-XRB consists of thermal blackbody component forming the softer region of the spectrum which is understood to be arising from the standard thin accretion disc that exists around the black hole \citep{1973A&A....24..337S}. Typical energy spectrum also consists of high energy non-thermal continuum that originates from the high energy plasma/corona that inverse-Comptonizes the soft radiation from the disc \citep{1980A&A....86..121S,1995ApJ...455..623C}. Along with these, additional features such as reflection hump at $\sim30$ keV and Fe emission line at $\sim6.4$ keV are also present in spectrum  which are the result of reflection of Comptonized radiation from the corona off the accretion disc \citep{1989MNRAS.238..729F,2003PhR...377..389R}. Depending on contribution from each of these components, different spectral states have been defined for the BH-XRBs (\citealt{1995ApJ...455..623C,2006csxs.book..157M,2012A&A...542A..56N,2021MNRAS.508.2447B,2022MNRAS.510.3019A,2023MNRAS.tmp..141P} and references therein). Such typical spectral states are observed in two extragalactic BH-XRBs LMC X-1 and LMC X-3 in which most of the time, the X-ray spectrum is dominated by thermal disc emission (see \citealt{2021MNRAS.501.5457B,2022AdSpR..69..483B}). However, currently, a comprehensive understanding of spectral states in the other three extragalactic BH-XRBs i.e., M33 X-7, NGC 300 X-1 and IC 10 X-1 does not exist. As a result, there is a lack of understanding of the geometry of their accretion discs. Therefore, in this work, we investigate the spectral properties of three extragalactic black hole binaries M33 X-7, IC 10 X-1 and NGC 300 X-1 in order to understand the accretion scenario and geometry by making use of all available observations with \textit{XMM-Newton} and \textit{NuSTAR}. General properties of each of these sources are as follows.

\subsection{M33 X-7}
M33 X-7 is located in the galaxy M33 at a distance of $840\pm20$ kpc \citep{2013ApJ...773...69G} from our Galaxy.  It is a highly inclined system with inclination angle $i=74.6^{\circ}\pm1.0^{\circ}$ \citep{2007Natur.449..872O}. The mass of BH is estimated to be $15.65\pm1.45$ M$_{\odot}$ from dynamical study which makes it one of the most massive stellar mass BHs ever found \citep{2007Natur.449..872O}. The companion star is an O super-giant of mass $\sim70$ M$_{\odot}$ which eclipses the BH. Such high mass of the companion facilitates the mass accretion by the BH via wind-Roche lobe overflow where the stellar wind is captured and fed to the inner Lagrangian point \citep{2022A&A...667A..77R} which enables the high luminosity emission (L $\sim10^{38}$ erg s$^{-1}$). The source is often seen in high luminosity state, where its energy spectrum is characterized by absorbed disc blackbody, Bremsstrahlung or slim-disc model \citep{2006ApJ...646..420P,2022A&A...667A..77R}. \cite{2008ApJ...679L..37L} has estimated the dimensionless spin parameter ($a$) of BH in this system to be $0.84\pm0.05$  from continuum-fitting method.
\subsection{NGC 300 X-1}
NGC 300 X-1 is a High Mass X-ray Binary (HMXB) system in which, the companion is a massive Woft-Rayet (WR) star, which makes the system belong to the rare WR+BH system \citep{2007A&A...461L...9C}. A WR star is in the final stage of its stellar evolution, experiences high mass loss through stellar wind. This wind gets accreted by the BH forming wind-Roche lobe overflow accretion disc, which emits X-rays of luminosity $\sim10^{36-39}$ erg s$^{-1}$ \citep{2021ApJ...910...74B}. It is located in the galaxy NGC 300 at a distance of $\sim2$ Mpc \citep{2009ApJS..183...67D}. \cite{2021ApJ...910...74B} measured the inclination angle of the system from the deep eclipses seen in its X-ray lightcurve to be within $60^{\circ}-75^{\circ}$ and binary period to be $\sim32.8$ hours. They estimate the mass of the BH to be $17\pm4$ M$_{\odot}$ using radial velocity measurement of C \RomanNumeralCaps 4 $\lambda$1550 emission line. Most often, the source energy spectrum is observed to be very steep with dominant non-thermal component and a thermal disc component imitating the Steep Powerlaw State (SPL) seen in Galactic BH-XRBs. However, it does not display the temporal properties generally associated with SPL \citep{2008A&A...488..697B}. Such spectral state is similar to that observed in several ULXs (\citealt{2011MNRAS.417..464M,2015ApJ...806...65W,2017ARA&A..55..303K}) even though the luminosity in this source always remains sub-Eddington.
\subsection{IC 10 X-1}
IC 10 X-1 is dynamically confirmed to be a WR+BH system similar to NGC 300 X-1, which implies that it is a HMXB source \citep{2004ApJ...601L..67B,2004A&A...414L..45C}. It is located in irregular dwarf galaxy IC 10 at a distance of $\sim715$ kpc \citep{2009ApJ...703..816K}. The orbital period and inclination angle of this source are measured to be $\sim34$ hours and $>63^{\circ}$, respectively \citep{2007ApJ...669L..21P}. Spectral state of the source is unidentified as spectral properties do not fall under any typically known spectral state \citep{2008A&A...488..697B}. The mass of BH is measure to be $23-33$ M$_{\odot}$ using radial velocity measurement of He \RomanNumeralCaps{2} $\lambda4686$ emission line \citep{2008ApJ...678L..17S}. If this mass is confirmed, IC 10 X-1 would be the most massive stellar mass BHs ever found. However, from the 5 hr long eclipse seen in its \textit{Chandra} and \textit{XMM-Newton} X-ray lightcurves, \cite{2015MNRAS.446.1399L} found a shift between He \RomanNumeralCaps{2} $\lambda4686$ line and mid-point of the eclipse. This would imply that the line originates from stellar wind or accretion stream rather than the star \citep{2023arXiv230213984B}. Thus, currently the mass of this BH remains unknown.

In this work, we carry out a comprehensive analysis of extragalactic BH-XRBs M33 X-7, NGC 300 X-1 and IC 10 X-1 using all available X-ray observations from \textit{XMM-Newton} \citep{2001A&A...365L..18S} and \textit{NuSTAR} \citep{2013ApJ...770..103H}. By carrying out the spectral modelling of the non-eclipse energy spectra, we infer the accretion geometry in these sources in their very high luminosity spectral states. Furthermore, since the previous mass estimate of these BHs are subjected to significant uncertainties, we derive their masses using continuum-fitting method, implementing the relativistic slim-disc model for the first time for these sources. We determine the size of eclipsing mass using eclipse period which we estimate by modelling the lightcurve. Eclipse properties are investigated by examining energy dependence of lightcurve, as well as by carrying out spectral analysis of eclipse spectrum.

This paper is organized as follows. In Section \ref{sec2}, we present the data and observations used for the analysis and method adapted to reduce the observational data. In Section \ref{sec4}, we present the spectral analysis and modelling procedures along with the results and in Section \ref{timing}, the timing and spectral analysis methods applied for studying eclipse data are discussed. Finally, in Section \ref{sec5}, we discuss the implication of obtained results in the context of accretion disc dynamics and infer accretion geometry in all five extragalactic stellar mass BH-XRBs i.e., M33 X-7, NGC 300 X-1, IC 10 X-1, LMC X-1 and LMC X-3.

\section{Observation and Data Reduction}
\label{sec2}
In this work, we make use of all observations of M33 X-7, NGC 300 X-1 and IC 10 X-1 carried out by \textit{XMM-Newton} and \textit{NuSTAR} till date. A log of considered observations within the period of 2000-2019 are listed in Table \ref{tab1}. Observations of M33 X-7 include pointed as well as off-axis observational data. In few of these observations, X-ray emissions are observed to undergo eclipses, resulting in a decrease in flux in their lightcurves. The period corresponding to ingress (eclipse entry), true eclipse (low flux phase) and egress (eclipse exit) estimated from visual inspection are mentioned in this table. We have made use of observations from \textit{XMM-Newton} (0.1$-$10 keV) and \textit{NuSTAR} (3$-$79 keV) since they can resolve the distant X-ray sources and avoid contamination from nearby sources. In addition, we have also made use of an \textit{XMM-Newton} data-set of LMC X-1 that is observed on 21-10-2000 with observation ID 0112900101.

\begin{table*}
\centering
\caption{Log table of observations of M33 X-7, NGC 300 X-1 and IC 10 X-1 by \textit{XMM-Newton} and \textit{NuSTAR} during the observational period of 2000-2019.} 
\begin{tabular}{ccccccccc}
\hline
Source      & Identifier & Observation ID & Observation Date & Observatory & \multicolumn{4}{c}{Exposure period (ks)}  \\ \cline{6-9}
     &  & & &  & non-eclipse  & Ingress & True eclipse & Egress \\
 \hline
 M33 X-7   & MX1  & 102640401    & 02-08-2000         & XMM-Newton  & 13.0 &  - &  - &  -  \\
      &  MX2 & 102641201   & 02-08-2000        & XMM-Newton  & 3.7 &  - &  -&  - \\
            & MX3  & 102640501  & 05-07-2001      & XMM-Newton  & 11.8 &  - &  -&  - \\
     & MX4   & 102642301      & 27-01-2002          & XMM-Newton  & 12.3                                                             &  - &  -&  - \\
           &  MX5 & 141980601 & 23-01-2003     & XMM-Newton  & 13.7 &  - &  -&  - \\
  &  MX6 & 141980801 & 12-02-2003    & XMM-Newton  & 10.2 &  -&  -&  -  \\   
   
    &  MX7 & 606370401 & 11-01-2010 & XMM-Newton  & 17.1  &  - &  -&  - \\  
     & MX8  & 606370701 & 21-01-2010  & XMM-Newton  & 15.6 &  - &  -&  - \\     
         & MX9  & 606371501 & 24-01-2010  & XMM-Newton  & 8.7  &  - &  -&  - \\      
           &  MX10 & 606370801 & 24-01-2010  & XMM-Newton  &  14.7 &  -&  -&  -  \\         
            &  MX11 & 606370901 & 28-01-2010  & XMM-Newton  & 19.7 &  -  &  -&  -\\         
             & MX12  & 606371001 & 31-01-2010  & XMM-Newton  & 14.7 &  - &  -&  - \\  
    &  MX13 & 606371201 & 07-02-2010  & XMM-Newton  & 18.6 &  -&  -&  -  \\    
 & MX14   & 650510201      & 11-07-2010        & XMM-Newton  & 71.8 &  - &   39.2 &  - \\
              & MX15  & 650510701 & 14-08-2010  & XMM-Newton  & 97.1 &  -&  -&  -  \\
                &  MX16 & 606370601 & 09-03-2011 & XMM-Newton  & 29.7 &  - &  -&  - \\
            & MN17     & 50310001002    & 04-03-2017         & NuSTAR      &  206.1                                                           & - &  -&  - \\
 \rowcolor{silver}       & MN18    & 50310001004    & 21-07-2017    & NuSTAR      & 206.4     &      -                                                  &  -&  - \\
      \rowcolor{silver}       & MX19     & 800350101      & 21-07-2017     & XMM-Newton  & 21.6                                                                  & -  &  -&  -  \\
            & MX20   & 831590401      & 13-07-2019         & XMM-Newton  &  19.6    & -                                                         &  -&  -\\
            & MX21    & 831590201      & 17-07-2019       & XMM-Newton  & 15.1 &                                                           - &  -&  - \\ \hline
NGC 300 X-1 & NX1   & 112800201      & 26-12-2000         & XMM-Newton  & 34.0 & -                                                            &  -&  -\\
            & NX2      & 112800101      & 01-01-2001     & XMM-Newton  & 30.8 &&  -&  13 \\
            & NX3    & 305860401      & 22-05-2005     & XMM-Newton  & 36.6  & -                                                          &  -&  - \\
            & NX4     & 305860301      & 25-11-2005     & XMM-Newton  & 36.6     & -                                                       &  -&  - \\
            & NX5     & 656780401      & 28-05-2010         & XMM-Newton  & 18.1  & -                                                          &  -&  - \\
    \rowcolor{silver}         & NN6$^{\dagger}$       & 30202035002    & 16-12-2016    & NuSTAR      & 328.2           & -                                                      &  -&  - \\
   \rowcolor{silver}          & NX7   & 791010101      & 17-12-2016       & XMM-Newton  & 137.4          & $20$                                                      & $10$ & $9$ - \\
            & NX8     & 791010301      & 19-12-2016        & XMM-Newton  & 35.6 & 21  &    19                                                      & 6  \\
            & NN9$^{\dagger}$      & 90401005002    & 31-01-2018        & NuSTAR      & 107.7   & -                                                           &  -&  - \\ \hline
IC 10 X-1   & IX1      & 152260101      & 03-07-2003     & XMM-Newton  & 28.3 & -& 4 & 13 \\
            & IX2    & 693390101      & 18-08-2012      & XMM-Newton  & 70.1 & 18  &  22 & 17                                                       \\
            & IN3$^{\ddagger}$    & 30001014002    & 06-11-2014      & NuSTAR      & 165.8  & -                                                         & -&- \\ \hline
\end{tabular}
\label{tab1}
\\ \raggedright
$^{\dagger}$ Obtained data is unusable since the source is too faint at high energies and due to the contamination from a nearby ULX source.\\
$^{\ddagger}$ The source is too weak at high energies, resulting in data to be statistically unreliable.
\end{table*} 

\begin{figure*}
\centering
\includegraphics[width=16cm,height=5cm,trim={0 13.6cm 0 0},clip]{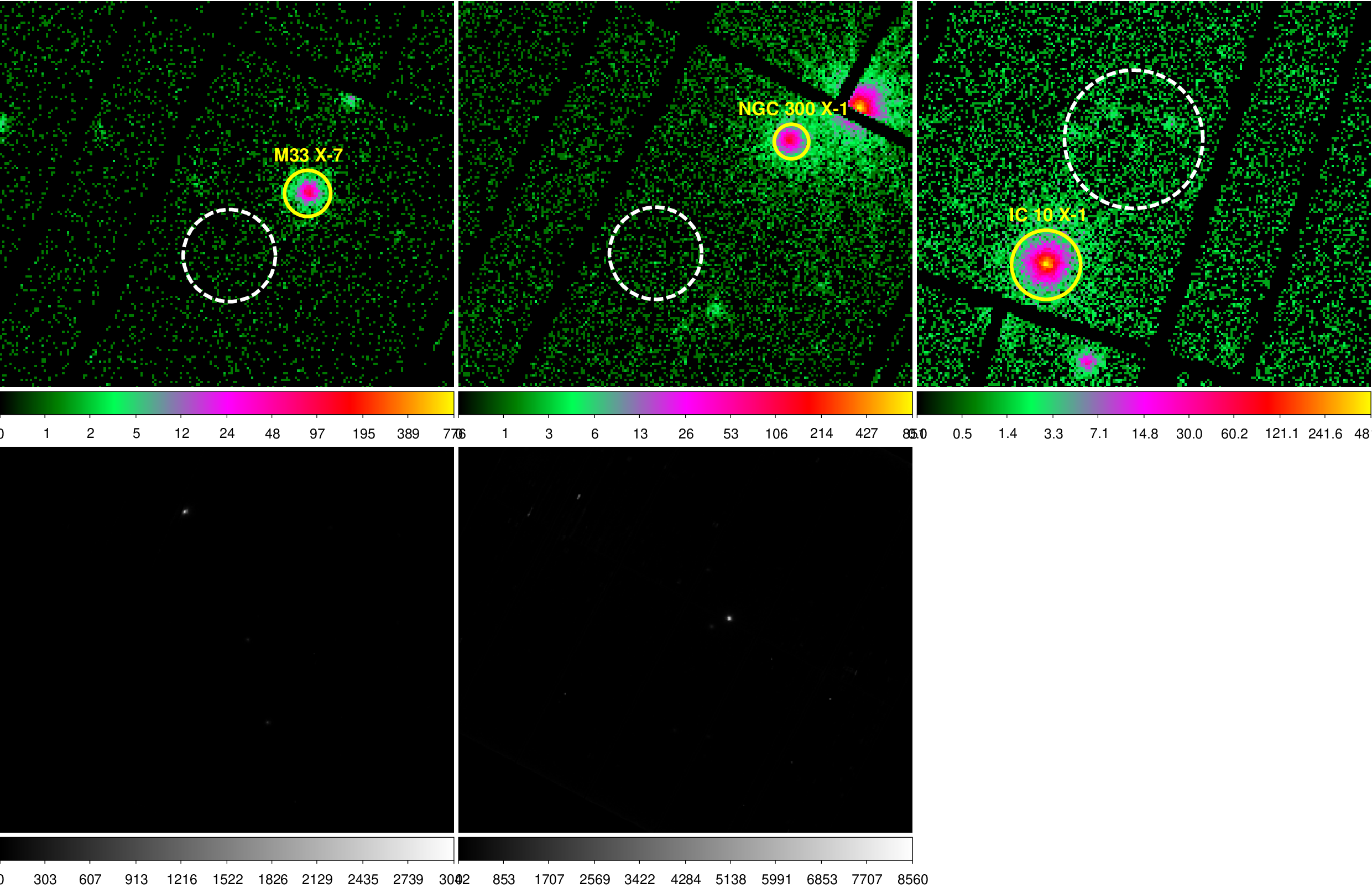}
\caption{ X-ray image of M33 X-7 (left panel), NGC 300 X-1 (middle panel) and IC 10 X-1  (right panel) obtained from \textit{EPIC-pn} instrument on-board \textit{XMM-Newton}. Images are filtered for background in the energy band 0.3$-$12 keV. Selected source region for the extraction of source spectrum and lightcurve are marked as circle in yellow color. A source free region with radius twice as that of the source region from which background spectrum is extracted is marked as circle in white color in all three panels. See the text for details.}
\label{ximage}
\end{figure*}

\subsection{XMM-Newton}
\label{sec2.1}
The \textit{EPIC-pn} and \textit{MOS} instruments on-board \textit{XMM-Newton} has observed M33 X-7 nineteen times, NGC 300 X-1 six times and IC 10 X-1 two times till date. For the reduction of these data, we use \texttt{HEASoft}\footnote{\url{https://heasarc.gsfc.nasa.gov/docs/software/heasoft/}} (version 6.30.1) and \texttt{Science Analysis System (SAS)}\footnote{\url{https://www.cosmos.esa.int/web/xmm-newton/home}} (version 19.1.0) softwares. The latest calibration files are used for the reduction. The \textit{SAS} command \texttt{emproc/epproc} is used to obtain science files from \textit{EPIC-MOS/pn}. To extract the source files, we use a circular region of radius 20$\arcsec$ for M33 X-7 and 30$\arcsec$ for IC 10 X-1.  In case of NGC 300 X-1, source files are extracted from circular region of 20$\arcsec$ radius when the nearby ULX NGC 300 ULX-1 was dim and $15\arcsec$ when it was bright to avoid contamination from it. Background files are extracted using larger circles from source free region. In Figure \ref{ximage}, we plot the background filtered X-ray images obtained in 0.3$-$12 keV of all three sources obtained using \textit{EPIC-pn} instrument on-board \textit{XMM-Newton}. Selected source regions are marked as yellow circles, and background regions as white circles.  \\

\subsection{NuSTAR}
\label{sec2.2}
\textit{NuSTAR} has two Focal Plane Modules i.e., \textit{FPMA} and \textit{FPMB} on-board, that observe the sources in broadband energy of 3$-$79 keV. It has observed the source M33 X-7 two times till date in which one of the observations (MN18) is simultaneous with an \textit{XMM-Newton} observation (MX19). NGC 300 X-1 is also observed twice by \textit{NuSTAR}, one of which is close to \textit{XMM-Newton} observation (NN6 and NX7). However, we find that the nearby ULX, NGC 300 ULX-1 \citep{2018MNRAS.479.3978K} is extremely bright in high energy while NGC 300 X-1 is very faint in these energies. Therefore, the emission from NGC 300 X-1 is completely overshadowed by the emission from the ULX. Hence, we do not consider any \textit{NuSTAR} observations of this source for our analysis. IC 10 X-1 has been observed only once by \textit{NuSTAR} which lies far from any of its \textit{XMM-Newton} observations and thus are analysed independently. We process these observational data using  \textit{NuSTAR} pipeline \texttt{NuSTARDAS}\footnote{\url{https://heasarc.gsfc.nasa.gov/docs/nustar/analysis/}} following standard reduction procedure. Clean event files are obtained using \texttt{nupipeline} and the source image is obtained using \texttt{XSELECT} from a circular region of radius 20$\arcsec$. Source spectra and lightcurves from both \textit{FPMA} and \textit{FPMB} are extracted from the selected regions using the command \texttt{nuproducts}.  


\section{Spectral Analysis and Results}
\label{sec4}
\begin{figure*}
\includegraphics[width=12cm,height=15cm,angle=-90]{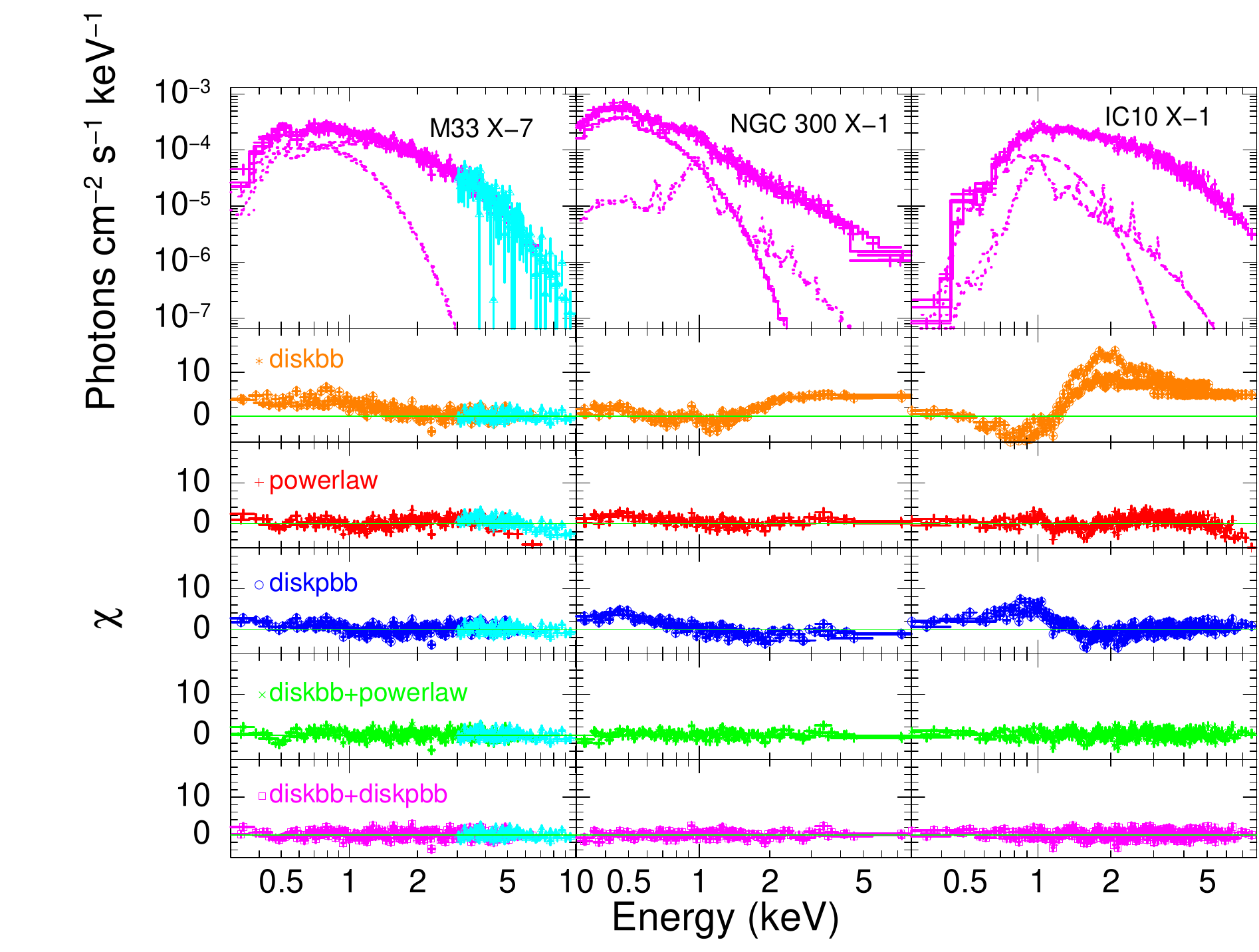}
\caption{The unfolded \textit{XMM-Newton} and \textit{NuSTAR} non-eclipse spectra of M33 X-7 (left), \textit{XMM-Newton} spectra of NGC 300 X-1 (middle) and IC 10 X-1 (right) that are modelled using Model-2 (\textit{Tbabs*Tbabs(diskbb+diskpbb+apec})) are plotted in the top panels. The \textit{XMM-Newton} spectra of all three sources comprised of \textit{EPIC-pn, MOS1} and \textit{MOS2} are plotted in magenta and the \textit{NuSTAR} spectra comprised of \textit{FPMA} and \textit{FPMB} of M33 X-7 are shown in cyan colour.  Spectra of M33 X-7, NGC 300 X-1 and IC 10 X-1 belong to Epoch MN18$+$MX19, NX4 and IX1 respectively. The variation of residuals of different model fits to these spectra i.e., \textit{diskbb}, \textit{powerlaw}, \textit{diskpbb}, \textit{diskbb+powerlaw} and \textit{diskbb+diskpbb} are illustrated by plotting  $\chi$ ((data-model)/error) of corresponding sources in the bottom panels. See the text for details.
 }
 \label{resi}
\end{figure*}

Spectral modelling of \textit{XMM-Newton} and \textit{NuSTAR} spectra belonging to all three sources are performed using \texttt{XSPEC v12.14.0} package of \texttt{HEASoft v6.33}.The spectrum extracted from \textit{MOS1}, \textit{MOS2} and \textit{pn} of \textit{XMM-Newton} belonging to each observation are modelled in the energy range of 0.3$-$8 keV to avoid the large error bars associated with high energy data points. All three of these spectra are modelled simultaneously in order to increase the statistics. For this, we use a cross normalization constant which is fixed to 1 in \textit{MOS1} data-set and allowed to vary freely in other two data-sets. Similarly, the \textit{FPMA} and \textit{FPMB} spectra from \textit{NuSTAR} are modelled together by adding a cross-normalization value which is fixed to 1 in \textit{FPMA} data-set, while allowed to vary in \textit{FPMB}. Fitting of \textit{NuSTAR} spectra is carried out in the energy range of 3$-$10 keV, 3$-$20 keV and 3$-$30 keV for M33 X-7, IC 10 X-1 and NGC 300 X-1 respectively since at higher energies, background dominates and no significant data is available. The spectra belonging to near simultaneous observations i.e., Epoch MN18 and MX19 as well as NN6 and NX7 are modelled jointly to have broadband spectral distribution.

All three sources exhibit eclipse in few of their \textit{XMM-Newton} lightcurves. Detailed spectral and timing analysis of such eclipse data is presented in Section \ref{timing}. In this section, we present the analysis and results of the non-eclipse observations that  excludes the data corresponding to the eclipse period. 
\begin{table*}
\caption{Best-fit parameters obtained by modelling the \textit{XMM-Newton} and \textit{NuSTAR} spectra of M33 X-7, NGC 300 X-1 and IC 10 X-1 using Model-1 (\textit{Tbabs*Tbabs(diskbb+apec+powerlaw)}). The flux is estimated in the energy range of 0.1$-$50 keV. Error of each parameter is estimated with 90\% confidence. See text for more details.}
\scalebox{0.8}{
\begin{tabular}{cccccccccccccc}
\hline
 &  &  Tbabs  & Tbabs                  & \multicolumn{2}{c}{diskbb}                                                                                           & powerlaw               &  \multicolumn{2}{c}{APEC}                                                                                         &                &                                                                                  &  &                                                & \\  \hline
 &     & \begin{tabular}[c]{@{}c@{}}$N_{H1}^{f}/10^{22}$\\ (cm$^{-2}$)\end{tabular}  & \begin{tabular}[c]{@{}c@{}}$N_{H2}/10^{22}$\\ (cm$^{-2}$)\end{tabular} & \begin{tabular}[c]{@{}c@{}}Tin\\ (keV)\end{tabular} & \begin{tabular}[c]{@{}c@{}}norm\\ $\times10^{-3}$\end{tabular} & $\Gamma$               & \begin{tabular}[c]{@{}c@{}}$kT$\\ (keV)\end{tabular} & \begin{tabular}[c]{@{}c@{}}$norm$\\ ($\times10^{-5}$)\end{tabular} & $\chi^{2}/dof$ & \begin{tabular}[c]{@{}c@{}}Flux ($\times10^{-12}$) \\ (erg cm$^{-2}$ s$^{-1}$) \\ (0.1-50 keV)\end{tabular}  &\begin{tabular}[c]{@{}c@{}}$f_{disc}$\\ (\%)\end{tabular} & \begin{tabular}[c]{@{}c@{}}$f_{pl}$\\ (\%)\end{tabular} & \begin{tabular}[c]{@{}c@{}}L \\ (L$_{Edd}$) \\ (0.1-50 keV)\end{tabular} \\ \hline
 
M33 X-7                                                & MX1 &  0.189 & $0.14^{+0.23}_{-0.14}$   & $0.94^{+0.15}_{-0.23}$ & $37.5^{+0.7}_{-0.9}$        &  $2.79^{+0.44}_{-0.26}$  & -   & - & 235.46/221 &  $5.65^{+0.69}_{-0.07}$     &   11 &  89 & $0.24\pm0.07$   \\ 
                                               & MX2 &  0.189 & $0.18^{+0.37}_{-0.14}$   & $0.72^{+0.13}_{-0.12}$ & $136^{+162}_{-66}$        &  $2.64^{+0.29}_{-0.19}$  & -   & - & 102.85/118 &  $4.42^{+0.48}_{-0.34}$     &   18 &  82 & $0.19^{+0.04}_{-0.05}$   \\ 
                                                                                              & MX3 &  0.189 & $0.08^{+0.17}_{-0.06}$   & $0.81^{+0.24}_{-0.25}$ & $51.5^{+11.3}_{-3.3}$        &  $2.36^{+0.37}_{-0.19}$  & -   & - & 165.77/192 &  $3.59^{+0.29}_{-0.68}$ &   12 &  88 & $0.15^{+0.04}_{-0.01}$    \\ 

& MX4 &  0.189 & $0.16^{+0.19}_{-0.12}$   & $1.12^{+0.09}_{-0.11}$ & $26.2^{+17.5}_{-9.9}$        &  $3.52^{+1.42}_{-0.92}$  & -   & - & 132.52/146 &  $6.00^{+0.13}_{-0.14}$     &   14 &  86 & $0.26^{+0.05}_{-0.04}$ \\ 

& MX5 &  0.189 & $0.05^{+0.04}_{-0.02}$   & $0.77^{+0.08}_{-0.09}$ & $122^{+67}_{-37}$        &  $2.55^{+0.45}_{-0.25}$  & -   & - & 187.12/207 &  $2.64^{+0.18}_{-0.19}$ &   34 &  66 & $0.11^{+0.03}_{-0.01}$  \\ 

& MX6 &  0.189 & $0.23^{+0.28}_{-0.17}$   & $0.96^{+0.10}_{-0.08}$ & $45.9^{+28.9}_{-20.6}$        &  $3.82^{+0.31}_{-0.29}$  & -   & - & 234.80/219 &  $4.45^{+0.24}_{-0.27}$ &   11 &  89 & $0.19\pm0.06$ \\ 

& MX7 &  0.189 & $0.05^{+0.16}_{-0.03}$   & $0.54^{+0.14}_{-0.08}$ & $393^{+478}_{-263}$        &  $1.89^{+0.24}_{-0.42}$  & -   & - & 227.81/196 &  $2.33^{+0.25}_{-0.20}$ &   30 & 70 & $0.09^{+0.06}_{-0.01}$ \\
& MX8 &  0.189 & $0.1^{+0.2}_{-0.1}$   & $0.85^{+0.41}_{-0.42}$ & $37.4^{+10.1}_{-35.7}$        &  $2.50^{+0.65}_{-0.36}$  & -   & - & 87.83/91 &  $3.78^{+0.44}_{-0.12}$ &   11 & 89 & $0.17^{+0.05}_{-0.02}$  \\
& MX9 &  0.189 & $0.05\pm0.04$   & - & -       &  $2.08^{+0.11}_{-0.10}$  & -   & - & 117.69/131 &  $1.12^{+0.32}_{-0.59}$ &   - & - & $0.11\pm0.01$  \\
& MX10 &  0.189 & $0.17\pm0.04$   & - & -       &  $2.52^{+0.09}_{-0.08}$  & -   & - & 125.92/184 &  $6.31^{+0.34}_{-0.23}$ &   - & - & $0.27^{+0.09}_{-0.07}$  \\
& MX11 &  0.189 & $0.37^{+0.20}_{-0.18}$   & $1.03^{+0.12}_{-0.10}$ & $50.0^{+33.3}_{-22.4}$     &  $2.54^{+0.30}_{-0.29}$  & -   & - & 205.18/228 &  $4.35^{+0.24}_{-0.21}$ &   21 & 79 & $0.19^{+0.04}_{-0.02}$  \\
& MX12 &  0.189 & $0.09^{+0.17}_{-0.09}$   & - & -   &  $2.29\pm0.07$  & -   & - & 184.43/197 &  $3.71^{+0.39}_{-0.21}$ &   - & - & $0.16^{+0.04}_{-0.02}$ \\
& MX13 &  0.189 & $0.23\pm0.04$   & - & -   &  $2.46\pm0.06$  & -   & - & 159.41/136 &  $3.14^{+0.27}_{-0.31}$ &   - & - & $0.13\pm0.02$ \\

& MX14 &    0.189 &    $0.07^{+0.07}_{-0.06}$    & $1.08\pm0.05$ & $14.6^{+3.8}_{-3.3}$ & $2.86^{+0.17}_{-0.16}$  & - & - & 327.35/245 & $2.88^{+0.03}_{-0.06}$ & 14 &   86 & $0.12\pm0.01$ \\

& MX15 &    0.189 &    $0.10^{+0.09}_{-0.08}$    & $1.10^{+0.18}_{-0.16}$ & $13.3^{+11.7}_{-8.0}$ & $2.79^{+0.28}_{-0.24}$  & - & - & 121.78/110 & $3.76^{+0.63}_{-0.87}$ & 18 &   82 & $0.16^{+0.03}_{-0.02}$ \\
& MX16 &    0.189 &    $0.10^{+0.27}_{-0.09}$    & $0.86^{+0.13}_{-0.18}$ & $57.5^{+60.0}_{-27.5}$ & $2.61^{+0.42}_{-0.26}$  & - & - & 108.58/134 & $1.29^{+0.23}_{-0.20}$ & 30 &   70 & $0.16^{+0.05}_{-0.02}$ \\

 & MN17 & 0.189  & $0.1^{f}$   & $0.90\pm0.08$                                       & $82.6^{+64.1}_{-35.4}$  & -                      &       -   & - &      110.53/105  &  $1.35\pm0.10$    & -   &      -   & $0.06\pm0.01$     \\
 
 \rowcolor{silver} & MN18$+$MX19 & 0.189  & $0.17^{+0.11}_{-0.08}$   & $0.98\pm0.04$ & $39.5^{+9.9}_{-8.4}$  &      $3.13^{+0.17}_{-0.15}$               & -  & - & 276.42/240       & $5.03\pm2.0$ & 10  &   90  & $0.22\pm0.02$ \\
 
 & MX20 & 0.189    & $0.11^{+0.12}_{-0.08}$      & $1.03\pm0.05$ & $36.1^{+9.3}_{-7.7}$       & $2.89^{+0.30}_{-0.25}$ &   -                   &         -                                                & 186.12/195     & $3.21^{+0.29}_{-0.76}$       & 27 &        73   & $0.14\pm0.02$ \\
 & MX21 & 0.189          & $0.16^{+0.14}_{-0.08}$ & $0.96\pm0.07$ & $43.4^{+13.2}_{-10.7}$  & $2.79^{+0.18}_{-0.15}$                    & -  &   -       & 195.60/180     & $4.67^{+1.68}_{-1.75}$    & 17  &   83 & $0.20\pm0.02$                                               \\ \hline
\begin{tabular}[c]{@{}c@{}}NGC\\  300 X-1\end{tabular} & NX1 & 0.09  & $<0.001$                & $0.14\pm0.01$                                       & $51.3^{+27.3}_{-16.4}$                                       & $1.97\pm0.14$          &  -                                                       &       -                                                  & 85.80/91       & $0.60^{+0.03}_{-0.05}$ &  54  & 46 & $0.14\pm0.01$ \\

& NX2 & 0.09 & $0.04\pm0.03$ & $1.39^{+0.27}_{-0.16}$                              & $2.44^{+1.8}_{-1.3}$                                           & $3.5^{+0.40}_{-0.38}$ & $0.96^{+0.02}_{-0.05}$  & $2.11\pm0.5$         & 160.41/160     & $4.3\pm0.3$ &  5             & 95                          &             $0.23^{+0.06}_{-0.05}$          \\
& NX3 & $0.09$ & $<0.002$         & $1.63^{+0.46}_{-0.26}$                              & $1.22^{+1.43}_{-0.80}$                                         & $3.32^{+0.22}_{-0.19}$ & $0.85\pm0.07$                                  & $1.89\pm0.40$                                  & 104.91/123     & $2.48\pm0.05$ & 8 & 91                                               & $0.64^{+0.34}_{-0.13}$ \\
& NX4 & 0.09 & $<0.02$          & $0.13\pm0.03$                                       & $103^{+307}_{-80}$                                       & $2.40\pm0.09$          &  $0.99^{+0.08}_{-0.07}$              &  $2.38\pm0.07$   & 180.00/150     & $1.69\pm0.09$  &  22           & 77                                              & $0.38^{+0.15}_{-0.08}$ \\

& NX5 & 0.09 &  $<0.02$                   & $0.18\pm0.02$                                                 & $21.8^{+13.9}_{-9.5}$  & $2.33\pm0.30$          & -                          & -                                 & 82.22/78       & $1.23^{+0.06}_{-0.09}$  &    31        &    66                                     & $0.24^{+0.04}_{-0.03}$ \\
 & NX7  & 0.09 & $0.03^{+0.03}_{-0.02}$                   &  $1.37\pm0.15$  & $2.70^{+1.4}_{-1.1}$  & $3.51^{0.51}_{-0.43}$ &   $0.99\pm0.05$ &       $6.74\pm0.01$                                                  & 143.37/150     & $1.54\pm0.04$  & 34        &      65 & $0.59\pm0.20$                                 \\
& NX8 & 0.09 & $0.06\pm0.04$          & $1.68^{+0.27}_{-0.20}$                              & $1.02^{+0.76}_{-0.51}$                                         & $3.92\pm0.50$ & $1.02\pm0.08$                                           & $1.82\pm0.05$                                            & 95.53/87     & $5.91^{+0.71}_{-0.64}$ &    34          & 65                  & $0.39\pm0.04$                             \\ \hline
IC 10 X-1                                              & IX1 & 0.5  & $0.42^{+0.15}_{-0.16}$         & $1.21\pm0.05$                                       & $49.1^{+12.1}_{-10.9}$                                         & $4.69^{+0.92}_{-1.08}$ & $0.95\pm0.07$                                           & $25.3^{+10.1}_{-8.1}$                                  & 314.97/275     & $18.0^{+0.28}_{-0.30}$ & 4.5                                                                           & 94   & $0.54\pm0.02$        \\
& IX2 & 0.5 & $0.51\pm0.09$ & $1.42\pm0.03$                                       & $17.8^{+2.3}_{-2.0}$                                           & $5.45^{+0.56}_{-0.53}$ & $0.96\pm0.03$                                           & $31.9^{+5.6}_{-4.9}$                                  & 449.98/353     & $24.0^{+0.13}_{-0.27}$ & 3.5  & 96                                  & $0.72\pm0.04$ \\ \hline  
\end{tabular}}
\\
\raggedright
\footnotesize{$^{f}$ Parameter frozen} 
\label{tab2}
\end{table*}

\begin{table*}
\caption{Best-fit parameters obtained by fitting Model-2 (\textit{Tbabs*Tbabs(diskbb+apec+diskpbb)}) to \textit{XMM-Newton} and \textit{NuSTAR} spectra of M33 X-7, NGC 300 X-1 and IC 10 X-1 during different observations. The flux is estimated in the energy range of 0.1$-$50 keV. Error of each parameter is estimated with 90\% confidence. See the text for more details. }
\scalebox{0.75}{
\begin{tabular}{ccclclcclcccccc}
\hline &         &       Tbabs            &  Tbabs        & \multicolumn{3}{c}{diskpbb}                                                                                                                   & \multicolumn{2}{c}{diskbb}                      & \multicolumn{2}{c}{APEC}                                              &        &        \\ \hline    &         & \begin{tabular}[c]{@{}c@{}}$N_{H1}^{f}$/$10^{22}$\\ (cm$^{-2}$)\end{tabular}  & \begin{tabular}[c]{@{}c@{}}$N_{H2}$/$10^{22}$\\ (cm$^{-2}$)\end{tabular} & \begin{tabular}[c]{@{}c@{}}$T_{in1}$\\ (keV)\end{tabular} & p                      & \begin{tabular}[c]{@{}c@{}}norm$\times10^{-3}$\\ \end{tabular} & \begin{tabular}[c]{@{}c@{}}$T_{in2}$\\ (keV)\end{tabular}            & norm                   & \begin{tabular}[c]{@{}c@{}}kT\\ (keV)\end{tabular} & \begin{tabular}[c]{@{}c@{}}norm\\ ($\times10^{-5}$)\end{tabular}  & $\chi^{2}/dof$  & \begin{tabular}[c]{@{}c@{}}$f_{dbb}$\\ (\%)\end{tabular}  &
\begin{tabular}[c]{@{}c@{}}$f_{dpb}$\\ (\%)\end{tabular} & \begin{tabular}[c]{@{}c@{}}$L^{\dagger}$\\ (L$_{Edd}$)\end{tabular}  \\ \hline
M33 X-7                                                & MX1     & 0.189                   & $0.08^{+0.02}_{-0.06}$            & $1.27^{+0.15}_{-0.19}$                                    & $0.50^{+0.05}_{\ddag}$          & $6.79^{+4.0}_{-2.6}$                                         & - & - &                      -      & -                        & 236.15/222   & - & - & $0.13^{+0.02}_{-0.01}$ \\ 
                                             & MX2     & 0.189                   & $0.08^{+0.02}_{-0.05}$            & $1.08^{+0.09}_{-0.08}$                                    & $0.50^{+0.02}_{\ddag}$          & $15.0^{+9.5}_{-1.9}$                                         & - & - &                      -      & -                        & 105.45/119  & - & - & $0.13^{+0.03}_{-0.01}$ \\ 
                                                                                          & MX3     & 0.189                   & $0.07^{+0.02}_{-0.07}$            & $1.38^{+0.18}_{-0.14}$                                    & $0.50^{+0.08}_{\ddag}$          & $5.89^{+3.8}_{-2.5}$                                         & - & - &                      -      & -                        & 169.98/193 & - & - & $0.11^{+0.02}_{-0.01}$   \\                
                                                                                                                       & MX4    & 0.189                   & $0.16^{+0.03}_{\ddag}$             & $1.17^{+0.21}_{-0.16}$                                    & $0.56^{+0.12}_{-0.05}$          & $14.0^{+3.0}_{-9.2}$                                         & $0.15\pm0.02$                  & $243^{+236}_{-106}$ &                      -      & -                        & 126.31/144     & 55 & 45  & $0.17\pm0.01$ \\ 
& MX5   & 0.189                   & $0.08\pm0.02$            & $1.10^{+0.08}_{-0.07}$                                    & $0.5^{+0.06}_{\ddag}$          & $11.8^{+5.5}_{-4.1}$                                         & -                    & -                &                      -      & -                        & 191.7/208  & - & -  & $0.11\pm0.01$  \\                                                                                           
                                                                                          & MX6   & 0.189                   & $0.05\pm0.03$            & $1.10^{+0.24}_{-0.18}$                                    & $0.51^{+0.02}_{-0.01}$          & $12.4^{+9.1}_{-5.6}$                                         & -  & -                      &                   -      &    -       & 234.90/220 & - & - & $0.09\pm0.01$ \\                                                                                     
                                                                                          
                                                                                  & MX7   & 0.189                   & $0.09^{+0.02}_{-0.04}$            & $1.28^{+0.21}_{-0.16}$                                    & $0.5^{+0.02}_{\ddag}$        & $6.54^{+4.6}_{-2.9}$                                         & -                    & -                &                      -      & -                        & 235.23/197   & - & - & $0.12\pm0.01$  \\                                                                                           
                                                                                  & MX8   & 0.189                   &$ 0.07^{+0.09}_{\ddag}$            & $1.24^{+0.21}_{-0.15}$                                    & $0.5^{+0.03}_{\ddag}$          & $6.86^{+5.9}_{-3.5}$                                         & -                    & -                &                      -      & -                        & 89.85/92 & - & - & $0.12\pm0.01$    \\                                                                                           
                                                                                  & MX9   & 0.189                   & $<0.01$            & $1.63^{+0.42}_{-0.41}$                                    & $0.53^{+0.04}_{\ddag}$          & $2.40^{+4.1}_{-2.1}$                                         & -                    & -                &                      -      & -                        & 123.18/132  & - & - & $0.10\pm0.01$  \\                                                                                           
                                                                                  & MX10   & 0.189                   & $0.09^{+0.02}_{-0.04}$   & $1.37^{+0.22}_{-0.17}$                                    & $0.5^{+0.03}_{\ddag}$          & $6.50^{+5.2}_{-3.2}$                                         & -                    & -                &                      -      & -                        & 128.89/185 & - & - & $0.14^{+0.02}_{-0.01}$   \\                                                                                           
                                                                                  & MX11   & 0.189                   & $0.10^{+0.02}_{-0.05}$            & $1.22^{+0.11}_{-0.09}$                                    & $0.5^{+0.04}_{\ddag}$          & $9.28^{+4.0}_{-2.9}$                                         & -                    & -                &                      -      & -                        & 209.45/229  & - & - & $0.14\pm0.01$  \\                                                                                           
                                                                                                                                                                    & MX12   & 0.189                   & $0.08^{+0.03}_{-0.07}$           & $1.34^{+0.18}_{-0.15}$                                    & $0.5^{+0.07}_{\ddag}$          & $5.84^{+3.8}_{-2.5}$                                         & -                    & -                &                      -      & -                        & 172.83/198  & - & - & $0.14^{+0.02}_{-0.01}$  \\                                                                                           
                                                                                                                                                                    & MX13  & 0.189                   & $0.09^{+0.03}_{-0.02}$            & $1.00^{+0.10}_{-0.09}$                                    & $0.5^{+0.02}_{\ddag}$          & $18.3^{+9.8}_{-6.6}$                                         & -                    & -                &                      -      & -                        & 133.90/137   & - & - & $0.12^{+0.02}_{-0.01}$ \\                                                                                           
& MX14     & 0.189                    &      $0.05^{+0.10}_{-0.05}$         & $1.18^{+0.12}_{-0.11}$                                    & $0.57^{+0.07}_{-0.04}$          & $6.92^{+7.2}_{-3.3}$                                         & $0.17\pm0.02$                  & $26.0^{+24.2}_{-10.3}$                 &                                             -  &-     & 323.07/244    &30 & 70 & $0.06^{+0.02}_{-0.01}$ \\
& MX15     & 0.189                    &      $<0.01$        & $1.38^{+0.25}_{-0.17}$                                    & $0.51^{+0.01}_{\ddag}$          & $3.34^{+3.25}_{-1.76}$                                         & -                 & - &                                             -  &-     & 119.04/111     & - & - & $0.07\pm0.01$ \\
& MX16  & 0.189                   & $0.03^{+0.06}_{-0.03}$            & $1.08^{+0.15}_{-0.11}$                                    & $0.55^{+0.03}_{-0.02}$          & $18.0^{+16.1}_{-10.1}$                                         & -                    & -                &                      -      & -                        & 106.23/135 & - & - & $0.10\pm0.01$   \\                                                                                           
& MN17     & 0.189                    & $0.1^{f}$            & $0.97\pm0.09$                                             & $0.58\ddag$                   & $32.4^{+26.0}_{-14.4}$                                         & -                      & -                      &                 -                             & -     & 110.27/104  & - & - & $0.06\pm0.01$   \\
\rowcolor{silver} & MN18+MX19 & 0.189                  & $0.18^{+0.11}_{-0.08}$          & $1.05^{+0.10}_{-0.09}$                                             & $0.58^{+0.09}_{-0.05}$          & $22.0^{+25.9}_{-11.0}$     & $0.15\pm0.01$                      & $218^{+133}_{-76}$                     &                                              - & -     & 262.67/239     & 46 & 54 & $0.10\pm0.01$ \\
& MX20     & $0.189$          & $<0.01$              & $1.17^{+0.10}_{-0.06}$                                    & $0.58^{+0.02}_{-0.03}$ & $14.1^{+3.9}_{-5.9}$                                           & -                  & - &     -                               & -                & 184.19/193     & - & - & $0.06\pm0.01$ \\
& MX21     & 0.189                   & $0.05\pm0.04$  & $1.16^{+0.11}_{-0.09}$                                    & $0.56^{+0.04}_{-0.03}$ & $14.4^{+11.1}_{-6.5}$                                          & -              & - &          -   &-                                       & 179.72/181   & - & - & $0.08^{+0.02}_{-0.01}$ \\ \hline

\begin{tabular}[c]{@{}c@{}}NGC\\  300 X-1\end{tabular} & NX1     & 0.09                 &      $<0.01$         & $1.72^{+0.32}_{-0.23}$    & $0.56^{+0.17}_{-0.06}$                     & $0.18^{+0.19}_{-0.10}$                                   & $0.15\pm0.01$ & $42.8^{+16.8}_{-11.8}$ & -  &-                                                & 83.30/91  & 67 & 35 & $0.12\pm0.01$   \\
& NX2     & 0.09                    &        $0.09^{+0.03}_{-0.03}$   & $1.60^{+0.19}_{-0.17}$  & $0.5^{+0.01}_{\ddag}$  & $55.2^{+0.51}_{-0.32}$                                  & $0.13\pm0.01$ & $102.9^{+106}_{-73.8}$   & $0.99^{+0.06}_{-0.05}$             & $2.39^{+0.04}_{-0.03}$               & 158.45/169 & 52 & 46  & $0.34^{+0.02}_{-0.01}$  \\
& NX3     & 0.09          &       $<0.01$       & $2.04^{+0.49}_{-0.51}$  & $0.5^{+0.02}_{\ddag}$                     & $0.18^{+0.69}_{-0.04}$   & $0.11\pm0.01$ & $271^{+288}_{-135}$   & $0.89^{+0.04}_{-0.22}$ & $0.10^{+0.13}_{-0.04}$                                                & 108.78/123 & 46 & 51 & $0.26\pm0.02$  \\
& NX4     & $0.09$                  &  $<0.01$             & $1.69^{+0.09}_{-0.21}$                                             & $0.5^{+0.01}_{\ddag}$                    & $0.44\pm0.05$                                  & $0.15\pm0.01$ & $58.7^{+41.5}_{-23.1}$   & $1.02^{+0.08}_{-0.06}$    & $2.38\pm0.05$                                  & 176.32/150 & 46 & 52 & $0.30\pm0.02$   \\
& NX5     & 0.09                  &    $<0.01$          & $1.56^{+0.74}_{-0.42}$  & $0.5^{+0.01}_{\ddag}$                     & $0.67^{+1.83}_{-0.64}$   & $0.20\pm0.02$ & $16.6^{+8.3}_{-5.3}$ & -                 & -          & 83.36/78   & 51	 & 49  & $0.21\pm0.02$  \\ 
& NX7    & 0.09                    &          $<0.01$     & $1.97^{+0.17}_{-0.23}$                       & $0.51^{+0.04}_{-0.02}$                 & $0.48^{+0.07}_{-0.06}$  & $0.14\pm0.01$ & $54.0^{+24.6}_{-13.8}$ & $1.02\pm0.04$ & $1.75\pm0.02$         & 148.97/149 & 49 & 48   & $0.28\pm0.01$  \\
& NX8     & 0.09                    &      $<0.01$         & $1.88^{+0.45}_{-0.44}$                                            & $0.53^{+0.07}_{\ddag}$                     & $0.21\pm0.09$                                         & $0.13\pm0.02$ & $54.2^{+81.8}_{-31.4}$ & $1.03\pm0.05$    & $3.10$                                  & 93.43/97  & 58 & 39 & $0.23\pm0.01$ \\ \hline
IC 10 X-1                                              & IX1     & 0.50                   &         $0.34^{+0.17}_{-0.15}$      & $1.26^{+0.10}_{-0.12}$                                             & $0.62^{+0.04}_{-0.03}$                     & $29.1^{+11.6}_{-6.9}$                                           & $0.19^{+0.11}_{-0.07}$          & $87.7^{+1.57}_{-0.77}$   & $1.03^{+0.09}_{-0.13}$   &  $0.13$ & 306.34/275  & 70 & 23 & $0.50\pm0.04$  \\
& IX2     & 0.5 &  $0.35^{+0.13}_{-0.12}$        & $1.41^{+0.06}_{-0.07}$             &         $0.66^{+0.06}_{-0.04}$              & $14.8^{+7.0}_{-4.5}$    & $0.15\pm0.01$ &  $276^{+95}_{-69}$                      & $1.03\pm0.05$                       & $18.5\pm3.3$               & 406.57/353 & 84 & 11 & $0.69\pm0.02$   \\ \hline
\end{tabular}}
\\
\raggedright
\footnotesize{$^{f}$ Parameter frozen} \\
\footnotesize{$^{\ddag}$ Parameter pegged at hard limit} \\
\footnotesize{$^{\dagger}$ For luminosity calculation of M33 X-7, NGC 300 X-1 and IC 10 X-1 source distance is considered to be 840$\pm20$ kpc, 2$\pm 0.03$ Mpc and 715$\pm50$ kpc  respectively and their respective mass to be $15.65\pm1.45$ M$_{\odot}$, 17$\pm4$ M$_{\odot}$ and 17$\pm4$ M$_{\odot}$. }

\label{tab3}
\end{table*}

\subsection{Modelling the Non-eclipse Spectra}
\label{sec4.1}

The spectral modelling of \textit{XMM-Newton} and \textit{NuSTAR} spectra is carried out using two absorption models (\textit{Tbabs}) along with the continuum fitting models. The hydrogen column density parameter ($N_{H1}$) of one of these models is fixed to the Galactic $N_{H}$ value by referring to \cite{2016A&A...594A.116H}, while the other ($N_{H2}$) is left to vary freely for all the \textit{XMM-Newton} fitting, which accounts for the interstellar absorption by the medium of host galaxy. Initially, we fit with the basic \textit{diskbb} and \textit{powerlaw} models to all the spectra by fixing $N_{H1}$ to $0.189\times10^{22}$ cm$^{-2}$, $0.09\times10^{22}$ cm$^{-2}$ and $0.5\times10^{22}$ cm$^{-2}$ for M33 X-7, NGC 300 X-1 and IC 10 X-1 respectively. Most of the \textit{XMM-Newton} fittings of NGC 300 X-1 and IC 10 X-1 are found to result in an emission residual at $\sim0.9$ keV. To account for this we include a thermal plasma emission model \textit{APEC} \citep{2001ApJ...556L..91S}. \textit{APEC} has plasma temperature parameter $kT$ that is left free, metal abundance parameter which is fixed to the default value of 1, and the normalization parameter that is free. Thus, we fit the spectra with the model combination \textit{Tbabs*Tbabs(diskbb+apec+powerlaw)}, which we refer to as Model-1. Parameters obtained from Model-1 are listed in Table \ref{tab2}. We estimate the unabsorbed source flux in 0.1$-$50 keV energy band using \textit{cflux} model.

In M33 X-7, we do not find $0.9$ keV emission line in any of the observations and hence in Model-1, we ignore \textit{APEC} and use \textit{diskbb} and \textit{powerlaw} models or \textit{powerlaw} alone where S/N is low. Model-1 results in photon index ($\Gamma$) of $1.89-3.82$ (see Table \ref{tab2}) and inner disc temperature ($kT_{in}$) of $\sim1$ keV in different observations. The value of $N_{H2}$, which is allowed to vary freely is found to be $<0.37\times10^{22}$ cm$^{-2}$ in different observations. It is fixed to $0.1\times10^{22}$ cm$^{-2}$ in epoch MN17 since it cannot be constrained due to unavailability of low energy data in \textit{NuSTAR} spectrum. The estimated unabsorbed source flux in 0.1$-$50 keV varies within the range $1.1-6.3\times10^{-12}$ erg cm$^{-2}$ s$^{-1}$ for this source. We find that the flux is dominated by non-thermal \textit{powerlaw} component with disc flux contribution ($f_{disc}$) $<34\%$ in all observations (see Table \ref{tab2}).

Model-1 fit to the NGC 300 X-1 spectra resulted in $N_{H2}<0.06\times10^{22}$ cm$^{-2}$ (see Table \ref{tab2}). Most of the bright observations are found to be associated with an emission peak at $\sim0.9$ keV, which is consistent with the previous observations \citep{2007A&A...461L...9C}. This emission line may corresponds to the emission from hot gas probably from the wind. The unabsorbed source flux is found to be within $0.6-5.9\times10^{-12}$ erg cm$^{-2}$ s$^{-1}$. Similar to the previous source, we observe that all spectra are generally steeper ($\Gamma=2.0-3.9$). It is observed that, in some of the Epochs, the disc blackbody exhibits peak values around $0.1-0.2$ keV, whereas in others, $kT_{in}$ is measured to be $1-2$ keV (see Table \ref{tab2}). Such change in disc temperature is not correlated with change in flux. Similar to M33 X-7, we find that the non-thermal component dominates the spectrum ($f_{disc}<34\%$) in majority of the observations except in Epoch NX1, where both components contribute almost equally. It is to be noted that NX1 has the minimum flux among all observations.

In case of IC 10 X-1, Model-1 fitting results in $N_{H2}$ value of $\sim0.42\times10^{22}$ cm$^{-2}$ and $\sim0.51\times10^{22}$ cm$^{-2}$ in Epoch IX1 and IX2 respectively. The $kT_{in}$ is close to 1 keV with a very steep index ($\Gamma=4.7-5.4$). Unabsorbed source flux is found to be high $1.8-2.4\times10^{-11}$ erg cm$^{-2}$ s$^{-1}$  during Epoch IX1 and IX2. Similar to other two sources, the non-thermal component dominate over thermal component ($f_{disc}<4.5\%$) in the spectrum.

These results suggest that the spectral behaviour of all three sources is similar. Steep spectral index ($\Gamma>2$) and high disc temperature ($kT_{in}\sim1$ keV) found in most of the observations indicate that sources are in soft spectral state. However, the dominant flux from powerlaw component in most of the observations suggests that it is unlikely to be in high/soft state. The observations with low $kT_{in}$ in NGC 300 X-1 cannot be categorized as the canonical hard spectral state for two main reasons: firstly, the spectra are considerably steeper compared to the typical hard state spectrum, and secondly, there is no significant correlation observed between the change in flux and $kT_{in}$, along with no noticeable change in flux between observations. The observed properties are rather similar to the  SPL state (see \citealt{2006csxs.book..157M} for details) seen in transient Galactic BH-XRBs during the high luminosity (L$>0.2$L$_{Edd}$) phase of its outburst. SPL state is interpreted as due to strong inverse-Comptonization of thermal radiation by corona. In order to investigate this, we replace \textit{powerlaw} model with Comptonization model such as \textit{simpl}, \textit{comptt} and \textit{nthcomp} in Model-1. Even though these models result in statistically acceptable fit, Comptonization parameters could not be constrained from any of these models.

Thus, we speculate that the dominance of the \textit{powerlaw} could be the result of extrapolation of \textit{powerlaw} component towards softer region of the spectrum. We illustrate this by plotting the residuals of fits with different models in Figure \ref{resi}. The residuals from the spectral fit using \textit{diskbb} alone (orange) shows excess residue in both softer ($<1$ keV) as well as higher ($>2$ keV) energies, which is accounted by the \textit{powerlaw} model when used along with \textit{diskbb} (green). This makes the Model-1 fit physically inappropriate as \textit{powerlaw} is meant to fit the Comptonized high energy continuum. Hence, we choose different modelling scenario, where the accretion disc acts like advection dominated slim-disc as the source luminosity approaches Eddignton limit (\citealt{1988ApJ...332..646A,2004MNRAS.353..980K,2017ApJ...836...48S}). 

In order to accomplish this, we model the spectra of all three sources using an absorbed slim disc model \textit{diskpbb}. The \textit{diskpbb} is a p-free model in which disc temperature is proportional to $r^{-p}$, where p is free parameter unlike \textit{diskbb} in which $T\propto r^{-0.75}$. While single \textit{diskpbb} yields good fit for most of the spectra of M33 X-7, some of them displayed soft excess at energies $<1$ keV, which we account for by using a standard \textit{diskbb} model. In order to confirm the need for an additional soft component, we carried out F-test which resulted in probability of $<10^{-20}$ in Epoch MX4, MX14 and MN18+MX19. In case of NGC 300 X-1, all energy spectra required additional soft component whose F-test resulted in probability of $<10^{-30}$. In IC 10 X-1, low energy \textit{diskbb} is required for Epoch IX1 and IX2 observations (F-test probability $<10^{-30}$). Thus, we fit the spectra with model combination \textit{Tbabs*Tbabs(diskbb+apec+diskpbb)}, which we refer to as Model-2. Estimated parameters from this model fit are listed in Table \ref{tab3}.

As stated above, the reason for choosing Model-2 is to explore the physical properties of the source in this peculiar spectral state. While Model-1 provides statistically satisfactory fits across all observations, in some cases, its reduced chi-square is comparable to that of Model-2, though in most observations, Model-2 offers slightly better fits. As a result, the residual variation in Model-2 is less pronounced than in Model-1. Additionally, Model-2 produces more tightly constrained parameters compared to Model-1.  While the difference may not be significant, Model-2 statistically provides a better fit than Model-1. However Model-1 hinders our ability to interpret the results correctly. Therefore, the primary motivation for considering Model-2 is not only its statistical significance but also Model-1’s inability to explain the underlying physical scenario.

Unlike Model-1, where \textit{diskbb} peaks at $\sim1$ keV, in Model-2, we see that \textit{diskbb} fits for the softer blackbody peak resulting in disc temperature $T_{in2}=0.1-0.2$ keV in all of the observations of all three sources. Steeper part of the high energy spectrum is now taken care by \textit{diskpbb} which results in higher disc temperature i.e., $T_{in1}\sim1$ keV in M33 X-7, 1.6$-$2.0 keV in NGC 300 X-1 and $1.32-1.4$ keV in IC 10 X-1. The p parameter is found to be $0.5-0.58$ in M33 X-7 and $0.5-5.56$ in NGC 300 X-1. In IC 10 X-1, p value is constrained to be $0.62^{+0.04}_{-0.03}$ and $0.66^{+0.06}_{-0.04}$ in Epoch IX1 and IX2 respectively. This implies the presence of hot, slim disc at the inner accretion region of these BHs which is represented by \textit{diskpbb}, along with a cooler standard accretion disc, which is represented by \textit{diskbb}. The flux contribution (see Table \ref{tab3}) from each disc component is calculated in 0.1$-$50 keV energy band, which shows that in M33 X-7 and NGC 300 X-1, \textit{diskpbb} contribution ($f_{dpb}$) is either equals or dominates the \textit{diskbb} fraction ($f_{dbb}$). In IC 10 X-1, \textit{diskbb} component dominates in both Epoch IX1 and IX2. The source luminosity in the units of Eddington luminosity listed in Table \ref{tab3} is calculated by considering distance to the source M33 X-7, NGC 300 X-1 and IC 10 X-1 to be $ 840\pm20$ kpc, $2\pm0.03$ Mpc and $715\pm50$ kpc respectively. In this calculation, we consider the black hole mass as reported in the literature from dynamical studies (see Section \ref{sec1}) i.e., $15.65\pm1.45$ M$_{\odot}$ and $17\pm4$ M$_{\odot}$ for M33 X-7 and NGC 300 X-1 respectively. In case of IC 10 X-1, since dynamical mass estimate is unavailable, we make an assumption that its mass is similar to that of NGC 300 X-1, based on the fact that both these sources are observed to show similar properties in spectral and timing aspects. A variation in luminosity is observed between the values obtained from Model-1 and Model-2. This variation primarily stems from differences in the $N_{H2}$ values obtained in these models at lower energies. The \textit{diskpbb} model can essentially extends only upto energies till 20 keV whereas the \textit{powerlaw} model can extend to higher energies, potentially leading to differences in the resulting luminosities at high energies. However, in all our observations, the data does not exceed 10 keV, so this would not have contributed to the observed change in luminosity. Nevertheless, we rely only on the luminosity estimate from Model-2, as the $N_{H2}$ value obtained from Model-2 is more dependable than that from Model-1 due to the ability of \textit{powerlaw} to modify the soft energies.

\subsection{Estimation of BH Mass}
\label{sec4.3}

We attempt to estimate mass of the BHs initially using normalization value of \textit{diskpbb} model obtained from Model-2 using its relation with radius of inner accretion disc ($R_{in}$). The $R_{in}$ is calculated using the relation, 
\begin{equation}  R_{in} \approx \xi \kappa^{2} (norm)^{1/2} (cos i)^{-1/2} D_{10} \text{ km},  \end{equation}
where $\kappa$ and $\xi$ are the correction factors i.e., $\kappa$ corresponds to color correction/hardening factor and $\xi$ corresponds to geometric factor.  We take $\kappa=1.7$ following \cite{1998PASJ...50..667K} for sub-Eddington sources. The value of $\xi$ is taken to be 0.353 for the slim-disc following \cite{2008PASJ...60..653V}. $D_{10}$, source distance in 10 kpc is considered to be 84, 200 and 71.5 for M33 X-7, NGC 300 X-1 and IC 10 X-1 respectively. Inclination angle ($i$) of M33 X-7 is considered to be $75^{\circ}$, NGC 300 X-1 to be $60^{\circ}-75^{\circ}$ and that of IC 10 X-1 to be 63$^{\circ}$. Then, by assuming that disc extends all the way till marginally stable orbit, we estimate the BH mass using the relation $R_{in} \approx \alpha GM_{BH}/c^{2}$.  To estimate the upper limit of the mass, we considered BH to be maximally spinning with $\alpha=1.24$  \citep{1972ApJ...178..347B}. Lower limit to the mass is estimated by considering the minimum spin estimated for these BHs. For M33 X-7 and IC 10 X-1, value of $a$ is known to be $>0.8$ \citep{2008ApJ...679L..37L,2016ApJ...817..154S}, which we consider to be lower limit. Although no spin estimate of NGC 300 X-1 has been performed to date, we assume it has a similar spin since its properties closely resemble those of M33 X-7 and IC 10 X-1. This allowed us to calculate the mass of M33 X-7 to be within 7.3$-$51 M$_{\odot}$, NGC 300 X-1 to be 1.3$-$27 M$_{\odot}$ and IC 10 X-1 to be within 3$-$30 M$_{\odot}$. 

\begin{figure*}
\includegraphics[width=10cm,trim={2cm 2.5cm 1cm 1cm},clip]{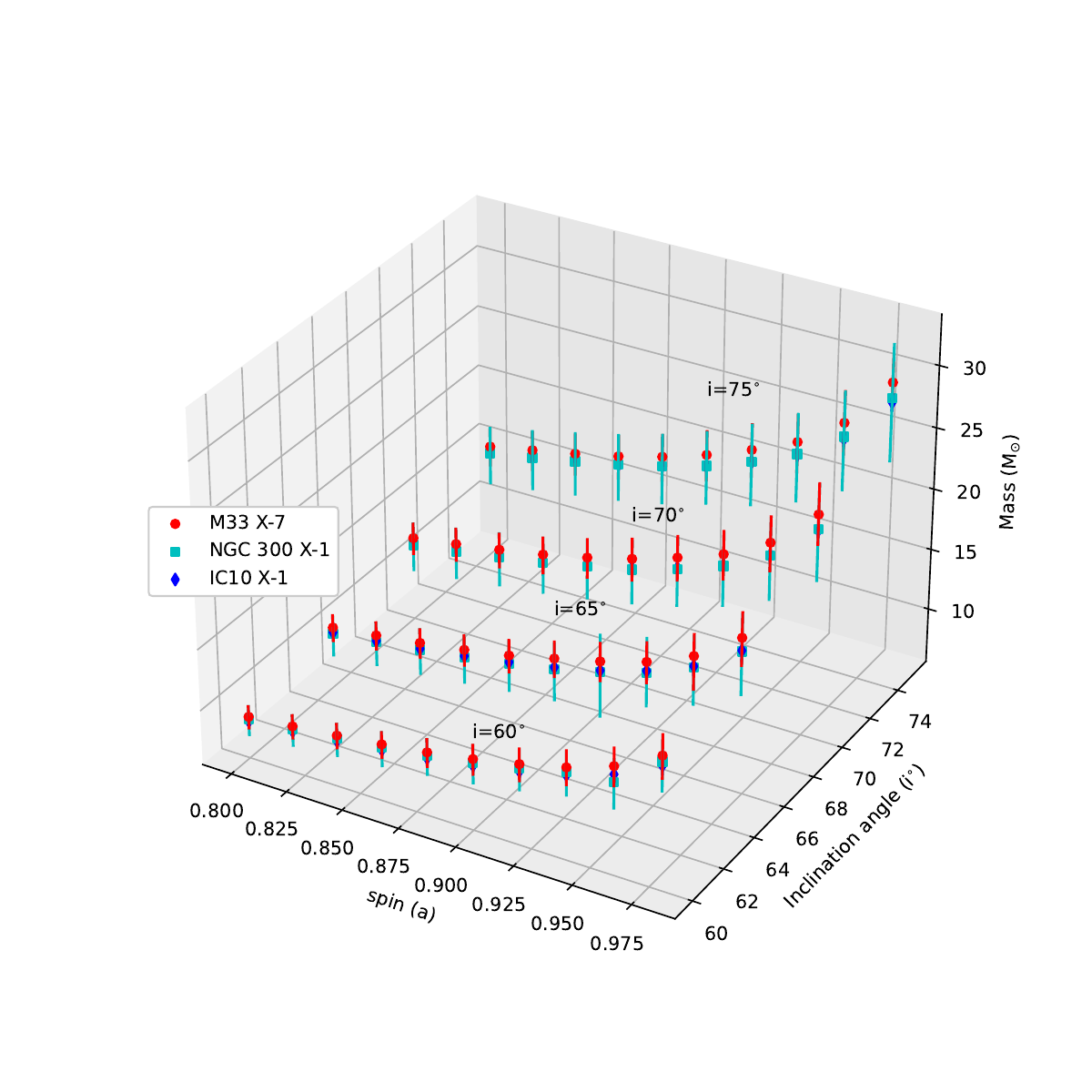}
\caption{Mass values of M33 X-7 (red), NGC 300 X-1 (cyan) and IC 10 X-1 (blue) obtained by fitting Model-3 (\textit{Tbabs*Tbabs(diskbb+apec+slimbh})). Spectral modelling is carried out with dimensionless spin (a) parameter fixed to $0.84$ for M33 X-7 and varied between 0.8 and 0.98 in steps of 0.02 for both NGC 300 X-1 and IC 10 X-1. Inclination angle is considered to be at four different values i.e., 60$^{\circ}$, 65$^{\circ}$, 70$^{\circ}$ and 75$^{\circ}$ for M33 X-7 and NGC 300 X-1 and 63$^{\circ}$, 65$^{\circ}$, 70$^{\circ}$ and 75$^{\circ}$ for IC 10 X-1. See the text for more details.}
\label{fig3}
\end{figure*}

It is evident that these mass ranges derived from the disc normalization value are very broad. This is due to the uncertainties associated with the normalization value of disc model which poses a challenge to determine the massive nature of these BHs. Uncertainty in disc normalization exists since \textit{diskpbb} model does not account for all the physical processes such as relativistic effects on disc emission. Therefore, in our subsequent step, we aim to derive better constraint of BH mass using continuum-fitting method by modelling the energy spectra using relativistic accretion disc model. In this approach, we use relativistic slim disc model \textit{slimbh} \citep{2011arXiv1108.0396S} in place of \textit{diskpbb} in Model-2. Model \textit{slimbh} is a relativistic model that allows to fit for the high luminosity (L $>0.3$ L$_{Edd}$) observations, where generally standard relativistic accretion disc model is invalid. Thus, we model the \textit{XMM-Newton} energy spectra of all three sources in 0.3$-$8 keV energy band using the model combination \textit{Tbabs*Tbabs(diskbb+apec+slimbh)}, which we refer to as Model-3. Here, \textit{diskbb} accounts for the standard cold disc, while \textit{slimbh} fits for the `hot' slim accretion disc. The $N_{H1}$ and $N_{H2}$ are frozen to the same values obtained from preliminary modelling (see Section \ref{sec4.1}). Distance to source is fixed to 840 kpc, 2 Mpc and 715 kpc for M33 X-7, NGC 300 X-1 and IC 10 X-1 respectively. In order to estimate the mass, spin and inclination needs to be fixed. We first apply this modelling to M33 X-7 spectrum since its inclination angle ($i=74.6\pm1.0^{\circ}$) and spin ($a=0.84\pm0.05$) are available from the literature (\citealt{2007Natur.449..872O,2008ApJ...679L..37L}). The source also has dynamically estimated mass value (M$=15.65\pm1.45$ M$_{\odot}$, \citealt{2007Natur.449..872O}). Therefore, by applying \textit{slimbh} modelling using the existing $i$ and $a$, we can verify the consistency of adapted mass estimation method. Thus, we carry out Model-3 fitting for MN18$+$MX19 observation and obtain the mass to be $14.86^{+1.44}_{-1.40}$ M$_{\odot}$. The obtained value is consistent with the dynamically estimated value, thus verifies the validity of mass estimation. However, the inclination angle of the inner accretion disc may not necessarily align with the orbital inclination angle of the binary system. Thus, to estimate the mass of the black hole considering its uncertainty, we must account for possible inclination angles, considering a misalignment of 10$-$15$^{\circ}$. Therefore, we carried out \textit{slimbh} modelling considering inclination angle with 60$-$75$^{\circ}$ varying in steps of 5$^{\circ}$ in order to explore variation in inclination angle of the source while keeping the spin fixed at 0.84. 

We then apply this modelling to NGC 300 X-1 and IC 10 X-1 by considering range of values for spin and inclination, as its accurate value is unknown. We assume both these sources to have spin $>0.8$, given that they are wind-fed systems \citep{2016ApJ...817..154S}. Such consideration of high spin in these systems are valid since HMXBs are likely to have high spin due to natal spin transfer from the massive progenitor star. Considering these systems have massive, young companion stars, this explanation is plausible \citep{2019ApJ...870L..18Q}. We freeze $a$ at various values ranging from 0.8 to 0.98 in steps of 0.02. The lower and upper limit for inclination angle of NGC 300 X-1 is known to be 60$^{\circ} $ and 75$^{\circ}$ respectively \citep{2021ApJ...910...74B}. In case of IC 10 X-1, inclination angle is $>63^{\circ}$ \citep{2007ApJ...669L..21P}. Thus, Model-3 fitting is carried out by fixing $i$ at four different values i.e., 60$^{\circ}$, 65$^{\circ}$, 70$^{\circ}$ and 75$^{\circ}$ for NX3 observation of NGC 300 X-1 and 63$^{\circ}$, 65$^{\circ}$, 70$^{\circ}$ and 75$^{\circ}$ for IX1 observation of IC 10 X-1. The luminosity parameter (L) and  BH mass parameter ($M_{BH}$) in these fits are allowed to vary freely while normalization is frozen to 1. The obtained value of $M_{BH}$ for various combinations of $i$ and $a$ in Epoch MN18$+$MX19, NX3 and IX1 of three sources are depicted in Figure \ref{fig3}. It is apparent from the figure that increase in inclination angle corresponds to an increase in mass of respective black holes. We determine the range of mass for M33 X-7, NGC 300 X-1 and IC 10 X-1 to be 8.9$-$14.9 M$_{\odot}$, 8.7$-$28 M$_{\odot}$ and 10.2$-$30 M$_{\odot}$ respectively.  It is to be noted that the derived mass values are associated with the uncertainties of inclination angle, spin and distance of sources. We attempt to obtain better constraint on spin and inclination angle by carrying out reflection modelling using several models such a \textit{relxill}, \textit{ireflect} etc. However, these models did not yield good constraints on reflection parameters due to absence of high energy data in the energy spectrum.

\begin{figure*}
\includegraphics[width=8.8cm,height=9.5cm,trim={0 0 0 0}]{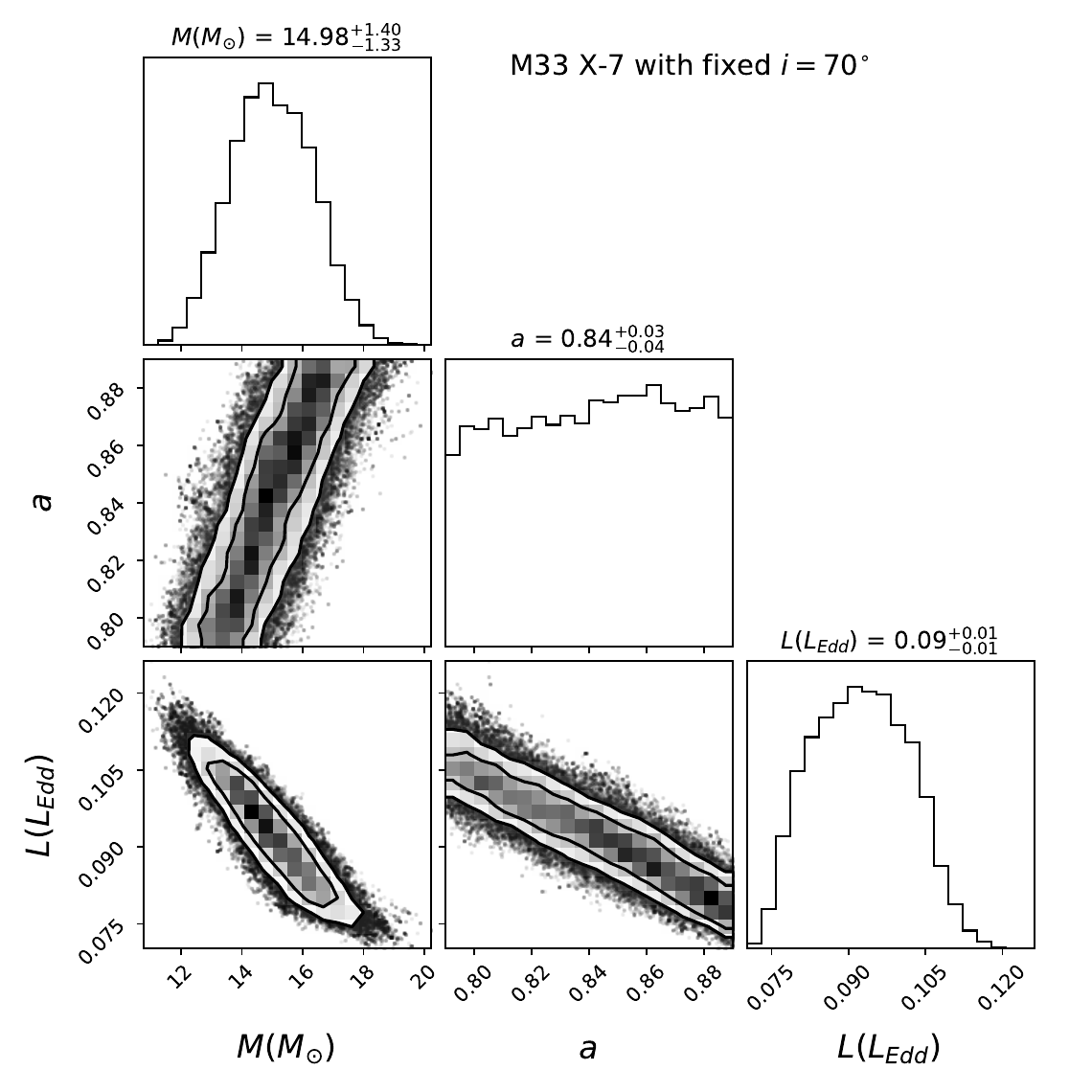}
\includegraphics[width=8.8cm,height=9.5cm,trim={0 0 0 0}]{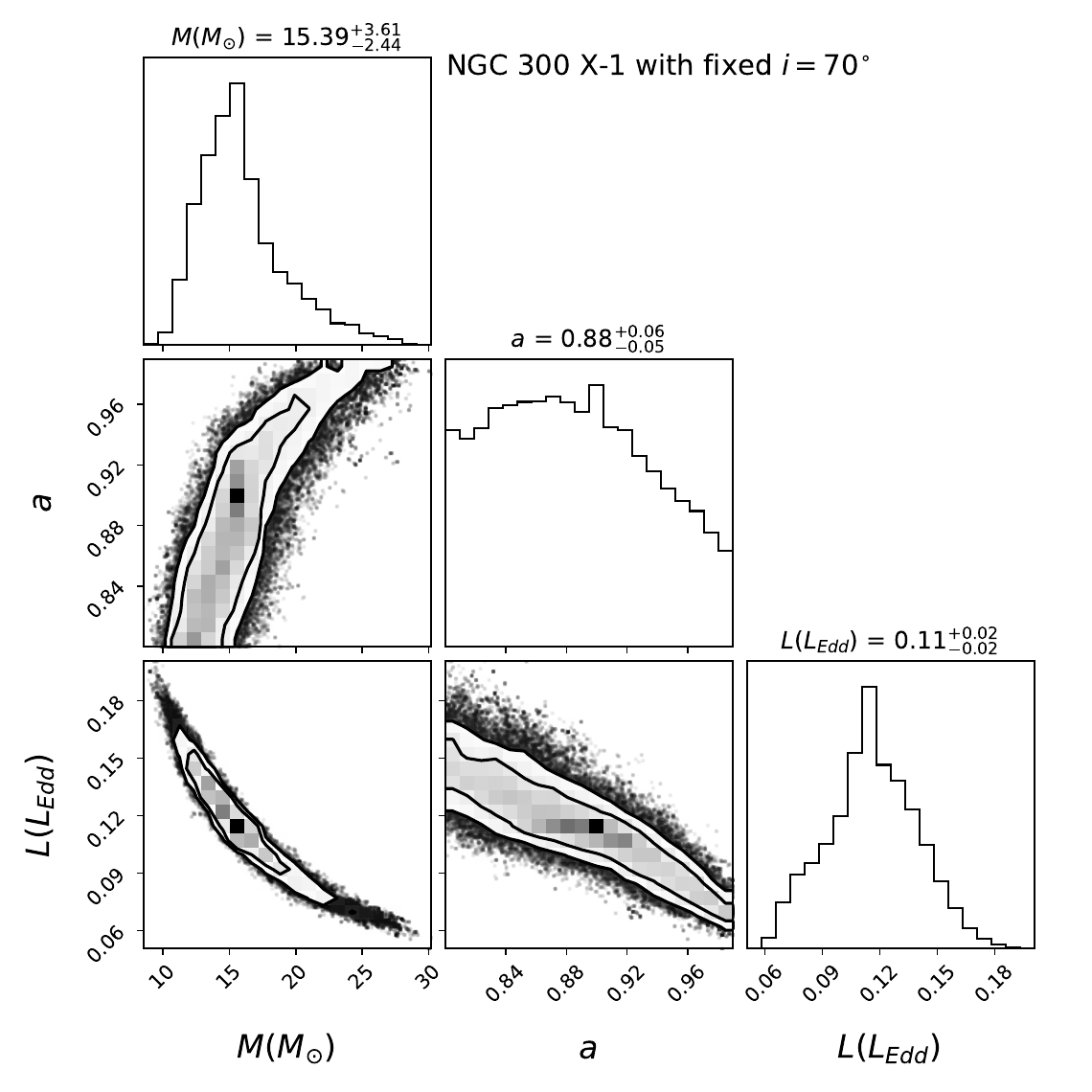}
\includegraphics[width=9cm,height=9.5cm,trim={0 0.5cm 0 0}]{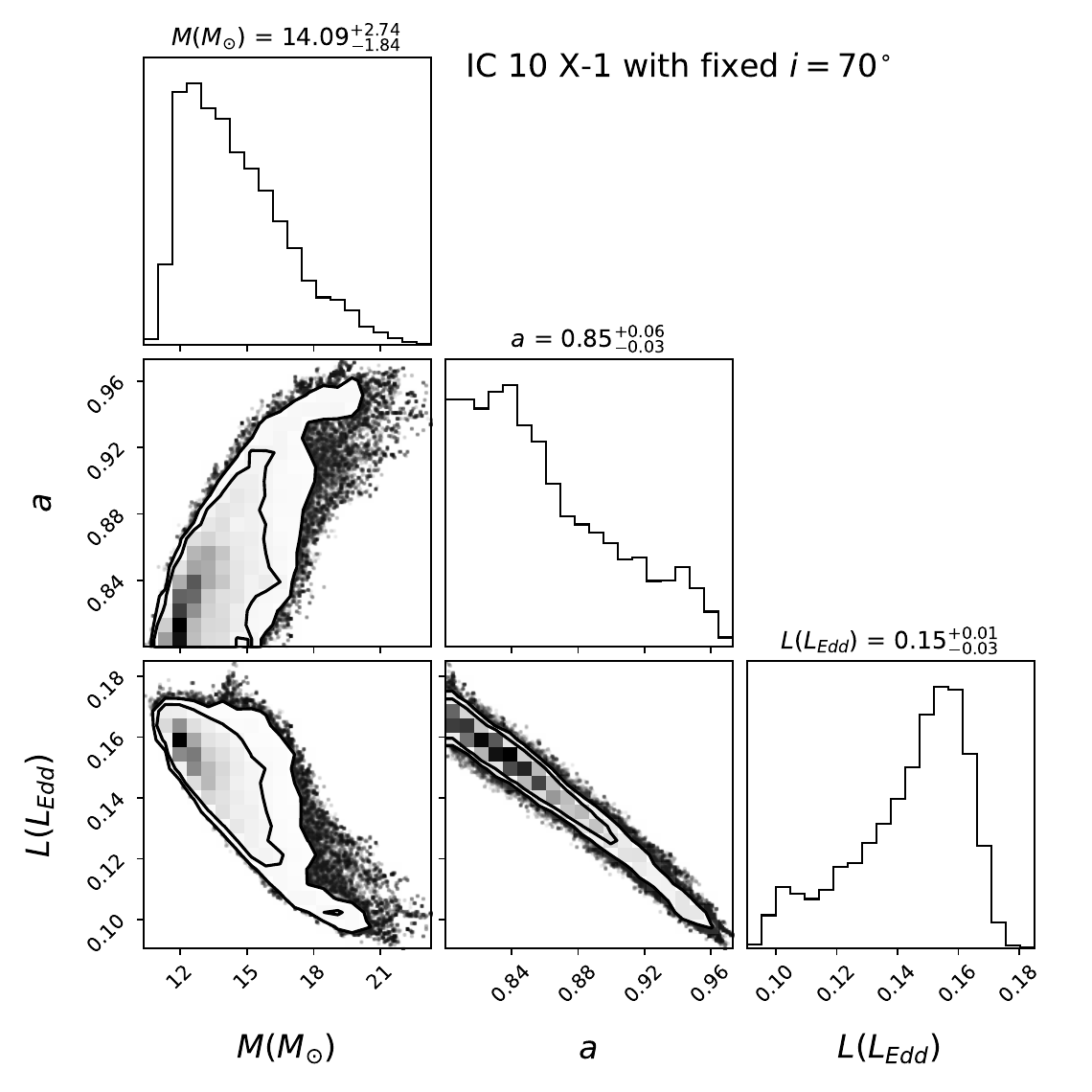}
\caption{Corner plot showing the variation of parameters of Model-3 fit i.e., luminosity, spin, and black hole mass for Epoch MN18$+$MX19 (top left), NX3 (top right) and IX1 (bottom) observations of M33 X-7, NGC 300 X-1 and IC 10 X-1 respectively. Marginalized distribution for each parameter is shown in histograms. Model-3 fit is carried out by keeping $M$ and $L$ free, $a$ constrained to vary within 0.8 to 0.98 for both NGC 300 X-1 and IC 10 X-1 and 0.79$-$0.89 for M33 X-7. $i$ is fixed to $70^{\circ}$. See the text for details.} 
\label{fig4}
\end{figure*}

We demonstrate the range of spin, luminosity and mass obtained within an observation of each source through corner plots in Figure \ref{fig4}. The fits are carried out by limiting the $a$ to vary within 0.8$-$0.98 for NGC 300 X-1 and IC 10 X-1 in order to explore the variation of mass within this spin range. However, if the $i$ is kept free or even limited to vary between 60$^{\circ}-70^{\circ}$ to explore the entire parameter range, the resultant fit is very poor. Hence we had to freeze the value of $i$. For representation purposes, we consider the fit performed with $i$ fixed at $70^{\circ}$ for all three sources. For M33 X-7, since the spin value is known to 0.84$\pm0.05$, simulation is carried for $a=$ 0.79$-$0.89 considering its uncertainty. To obtain these plots, we perform Markov Chain Monte Carlo (MCMC) simulations for Model-3 fits using Goodman-Weare algorithm \citep{2010CAMCS...5...65G} in \textit{Xspec} using \texttt{CHAIN} command. The chain is obtained for a total length of 200000, burn length of 50000, and  20 walkers. Obtained chain of parameter values from this simulation are then used to plot the corner plot using MCMC Hammer algorithm \citep{2013PASP..125..306F}. In Figure \ref{fig4}, we show the corner plots obtained for Epoch MN18$+$MX19, NX3 and IX1 observations. 

\begin{figure*}
\includegraphics[width=18cm,height=9.2cm]{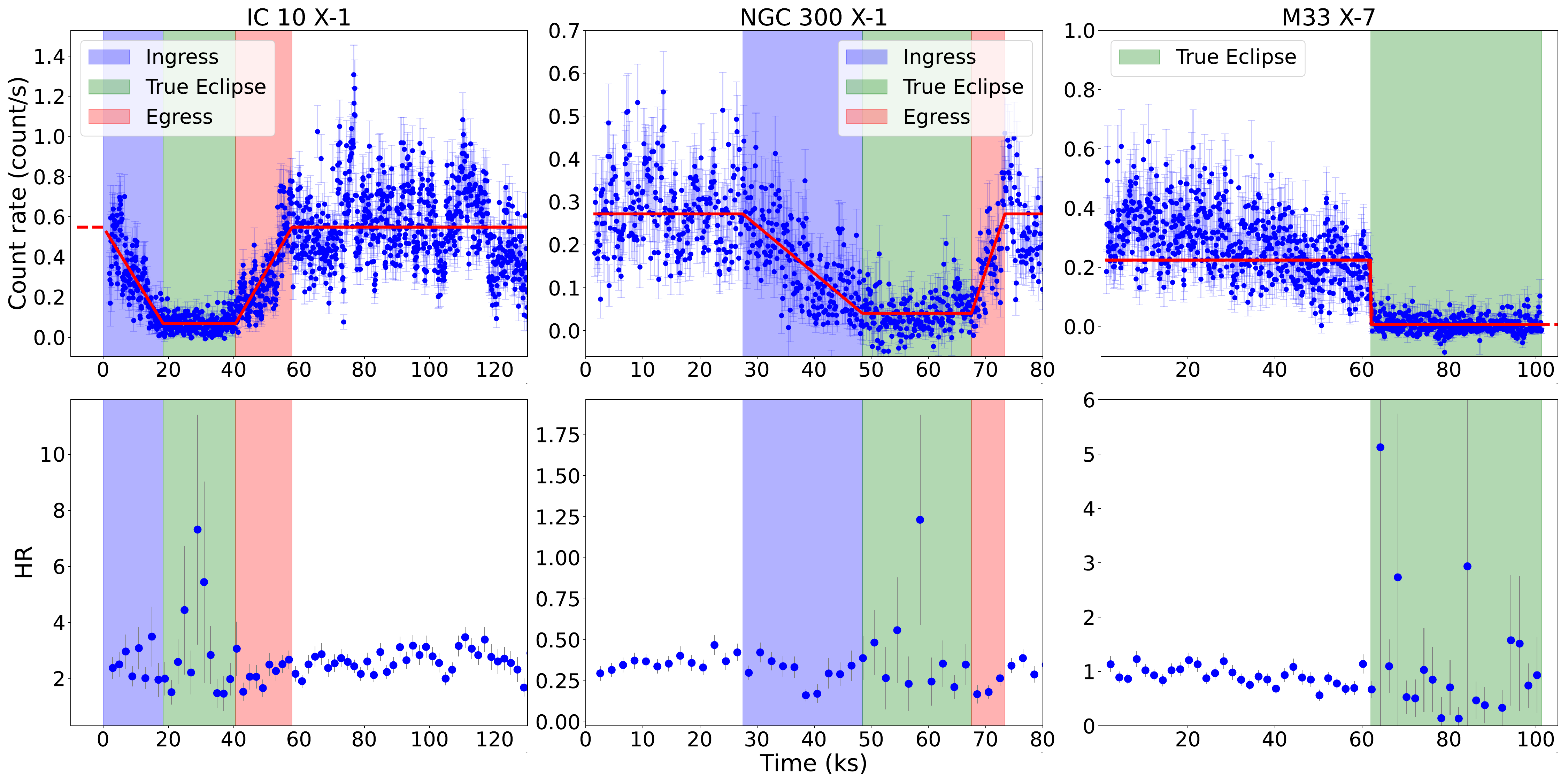}
\caption{\textit{EPIC-pn} lightcurves of IC 10 X-1, NGC 300 X-1 and M33 X-7 of Epoch IX2, NX8 and  MX14 extracted in the energy range of 0.3$-$8 keV are plotted with time-bin of 100 sec in top panels. Fitted model is plotted in red for all three lightcurves. Hardness ratio (2$-$8 keV/0.3$-$2 keV)  of these observations obtained using their corresponding hard and soft band lightcurves are plotted in bottom panels with time-bin of 2000 sec to provide better illustration of HR variations. See the text for details. }
\label{lc}
\end{figure*}
\section{Eclipse Data Analysis}

\label{timing}
IC 10 X-1, NGC 300 X-1 and M33 X-7 show eclipse in their \textit{XMM-Newton} lightcurve during the Epoch IX1, IX2, NX2, NX7, NX8 and MX14 (see Table \ref{tab1}). Out of these observations, complete eclipse is observed only in Epoch IX2 and NX8. In order to explore the nature and cause of eclipse, we analyse the lightcurve and energy spectrum during the eclipse phase in Epoch IX2, NX8 and MX14 and compare it with that of non-eclipse observations. The results of this eclipse analysis is presented in the following subsections.
\subsection{Lightcurve Analysis}
\label{lcan}
The eclipsed \textit{XMM-Newton} lightcurves of three sources during Epoch IX2, NX8 and  MX14 extracted in the total energy band of 0.3$-$8 keV are shown in top panel of Figure \ref{lc}. These lightcurves are plotted with bin-time of 100 sec. To estimate the duration of ingress, true eclipse and egress, we model the eclipse lightcurves using a mathematical function. Following \cite{2009ApJS..183..156W}, we use the ramp and step function to model the ingress, true eclipse and egress of the eclipse. In our model, we include five parameters i.e., out-of-eclipse flux, eclipse depth, duration of ingress, duration of true eclipse and duration of egress. The best-fit parameters are obtained by employing the Nelder-Mead optimization method \citep{10.1093/comjnl/7.4.308}. The functional form of the employed model is,

\begin{equation}
m(t) = b_{oe} - I_{\text{depth}} * \int R_{in}(t)R_{fl}(t)R_{eg}(t),
\end{equation}
 where, $b_{oe}$ is the out-of-eclipse flux and $I_{\text{depth}}$ is the depth of eclipse. The ramp up component is defined as, 
 \begin{equation}
 \label{eq3}
    R_{in}(t)= 
\begin{cases}
	0,& \text{if } t < 0 ,\\
    \frac{t}{t_{in}},& \text{if } 0\leq t < \text{ingress time $t_{in}$}, \\
    1,              & \text{if } t\geq \text{ingress time  $t_{in}$,}
\end{cases}
 \end{equation}
ramp down, $R_{eg}(t)$ is defined as to mirror this function, and the flat component, $R_{fl}={1, \text{for all t}}$.

We fit this model to IC 10 X-1 lightcurve, which shows symmetric eclipse in Epoch IX2, by allowing all five parameters to vary freely. The eclipse in NGC 300 X-1 lightcurve experiences a gradual entrance into eclipse followed by low flux phase and then a relatively swift exit. The spread of the initial data points corresponding to out-of-eclipse are wide and therefore while implementing the model, the out-of-eclipse flux value had to be fixed. In case of M33 X-7, we observe the ingress and true eclipse, however we do not observe the whole eclipse along with egress due to lack of observational data. The source appears to enter the eclipse with a sudden drop in counts. Therefore, we model it using only step function in Equation \ref{eq3} to determine the period of true eclipse using three parameters i.e., out-of-eclipse flux, duration of true eclipse, and the eclipse depth. We also attempted to model this lightcurve using slow ingress model instead of step model, given that the counts slowly starts to decline past 40 ks period in the lightcurve. However, identifying the end of ingress posed a challenge because of abrupt drop in flux just before the start of true eclipse. In top panel of Figure \ref{lc}, the fitted model is represented in red for all three lightcurves. In the IC 10 X-1 lightcurve, we extend model to the left, to illustrate the out-of-eclipse data before the eclipse period. In M33 X-7, we extend the true eclipse model for an additional duration of 4.54 ks to demonstrate the total eclipse based on the true eclipse period determined by \cite{2006ApJ...646..420P}. The extended model regions are plotted as dotted line. The resultant eclipse duration during ingress, true eclipse and egress are listed in Table \ref{tabec}. The egress, true eclipse and ingress period of IC 10 X-1 to be $19.0\pm0.2$ ks ($5.3\pm0.1$ hours), $18.4\pm0.4$ ks ($5.1\pm0.1$ hours) and $22.4\pm0.7$ ks ($6.2\pm0.1$ hours) respectively. Duration of egress, true eclipse and ingress in NGC 300 X-1 are determined to be $20.9\pm1.1$ ks ($5.8\pm0.3$ hours), $19.1\pm0.6$ ks ($5.3\pm0.2$ hours) and $5.9\pm0.4$ ks ($1.6\pm0.1$ hours) respectively. The minimum flux duration during Epoch MX14 of M33 X-7 is estimated to be $39.0\pm1.8$ ks ($10.8\pm0.5$ hours). 
\begin{table*}
\caption{Parameters obtained from the fitting of eclipse lightcurves. Duration of eclipses obtained by modelling the ramp and step function to the eclipse light curve. The uncertainties are estimated with 1$\sigma$ confidence.}
\begin{tabular}{|c|ccc|c|}
\hline
Source      & \multicolumn{3}{c|}{Duration (s)}                                                                          & \begin{tabular}[c]{@{}c@{}}Radius of companion star\\ (R$_{\odot}^{*}$)\end{tabular} \\ \hline
            & \multicolumn{1}{c|}{Ingress}              & \multicolumn{1}{c|}{True eclipse}         & Egress              &                                                                                  \\ \hline
M33 X-7     & \multicolumn{1}{c|}{-}                    & \multicolumn{1}{c|}{38988.42$\pm$1815.58} & -                   & 18.12$\pm0.98$                                                                   \\ \hline
NGC 300 X-1 & \multicolumn{1}{c|}{20924.95$\pm$1151.45} & \multicolumn{1}{c|}{19112.52$\pm$572.76}  & 5863.03$\pm$419.62  & 10.0$\pm0.87$                                                                    \\ \hline
IC 10 X-1   & \multicolumn{1}{c|}{19026.73$\pm$240.87}  & \multicolumn{1}{c|}{18436.16$\pm$400.04}  & 22383.38$\pm$711.55 & 9.99$\pm$0.74                                                                    \\ \hline
\end{tabular} \\
\raggedright
$^{*}$ $1R_{\odot}=695700$ km 
\label{tabec}
\end{table*}

Considering that eclipse is caused by the companion star, we then use the period of true eclipse to estimate its size following \cite{2015MNRAS.446.1399L}, using the relation, \begin{equation}
R_{cs} = T_{eclipse} \frac{(v1 + v2)}{ 2 sin(i)},
\end{equation}
where, $R_{cs}$ is the radius of companion star, $T_{eclipse}$ is the eclipse period corresponding the low flux/true eclipse, $v1$ and $v2$ are the tangential velocity of two objects in binary system. The velocities are estimated using the mass function relation (see \citealt{Carroll2007}). In this calculation, we assume mass of compact object to be the average black hole mass obtained in Section \ref{sec4.3}, while mass of the companion is sourced from the literature (see Section \ref{sec1}). Using the $T_{eclipse}$ obtained from the lightcurve modelling of Epoch IX2, NX8 and MX14, the obtained radii of the companion stars are listed in Table \ref{tabec}. The size of companion star of IC 10 X-1, NGC 300 X-1 and M33 X-7 are obtained to be 9.99$\pm$0.74 R$_{\odot}$, 10.00$\pm0.87$ R$_{\odot}$ and 18.12$\pm$0.98 R$_{\odot}$  respectively.

Further, we examine the energy dependence of the dips in flux by calculating the Hardness Ratio (HR) i.e., ratio of counts in hard energy band (2$-$8 keV) to soft energy band (0.3$-$2 keV). In the bottom panel of Figure \ref{lc}, we plot the variation of HR with time during Epoch IX2, NX8 and MX14 with bin-time of 2000 sec. These indicate an increase in HR variation during eclipse in IC 10 X-1 and NGC 300 X-1. In case of M33 X-7, even though we observe increase in HR, the associated error is large and falls within that of data points outside eclipse. To check the statistical significance of the change in HR during the eclipse, we estimate the percentage variation of mean count during eclipse with that of mean count out of eclipse. The percentage change in mean value of HR in true eclipse with respect to mean value of HR outside true eclipse is 25\% $\pm$ 45 \% in M33 X-7, 53 \% $\pm$ 35 \% in NGC 300 X-1 and 20 \% $\pm$ 21 \% in IC 10 X-1. This implies that variation in NGC 300 X-1 can be considered as significant while that in the other two sources cannot. In order to further explore the eclipse feature, we carry out comparative spectral study of eclipse and non-eclipse data, which is presented in Section \ref{4.2}.

To explore the variability, we also extract Power Density Spectrum (PDS) from the lightcurve of each observation. However, the PDS are mostly dominated by noise in the entire frequency range of 1 mHz $-$ 10 Hz and therefore we do not proceed further for PDS study.

\subsection{Modelling the Eclipse Spectra}
\label{4.2}

The energy spectra are extracted separately for true eclipse and non-eclipse segments of the Epoch IX2 and NX8. We model the in-eclipse and out-of-eclipse spectra of NGC 300 X-1 and IC 10 X-1 using Model-1 simultaneously but by allowing the parameters to vary independently. We add an additional partial covering model \textit{pcfabs} to Model-1 to account for absorption during eclipse whose parameters are  allowed to vary freely for eclipse spectra and frozen to 0 for out-of-eclipse spectra.  

\begin{figure*}
\includegraphics[height=13cm,width=10cm,angle=-90]{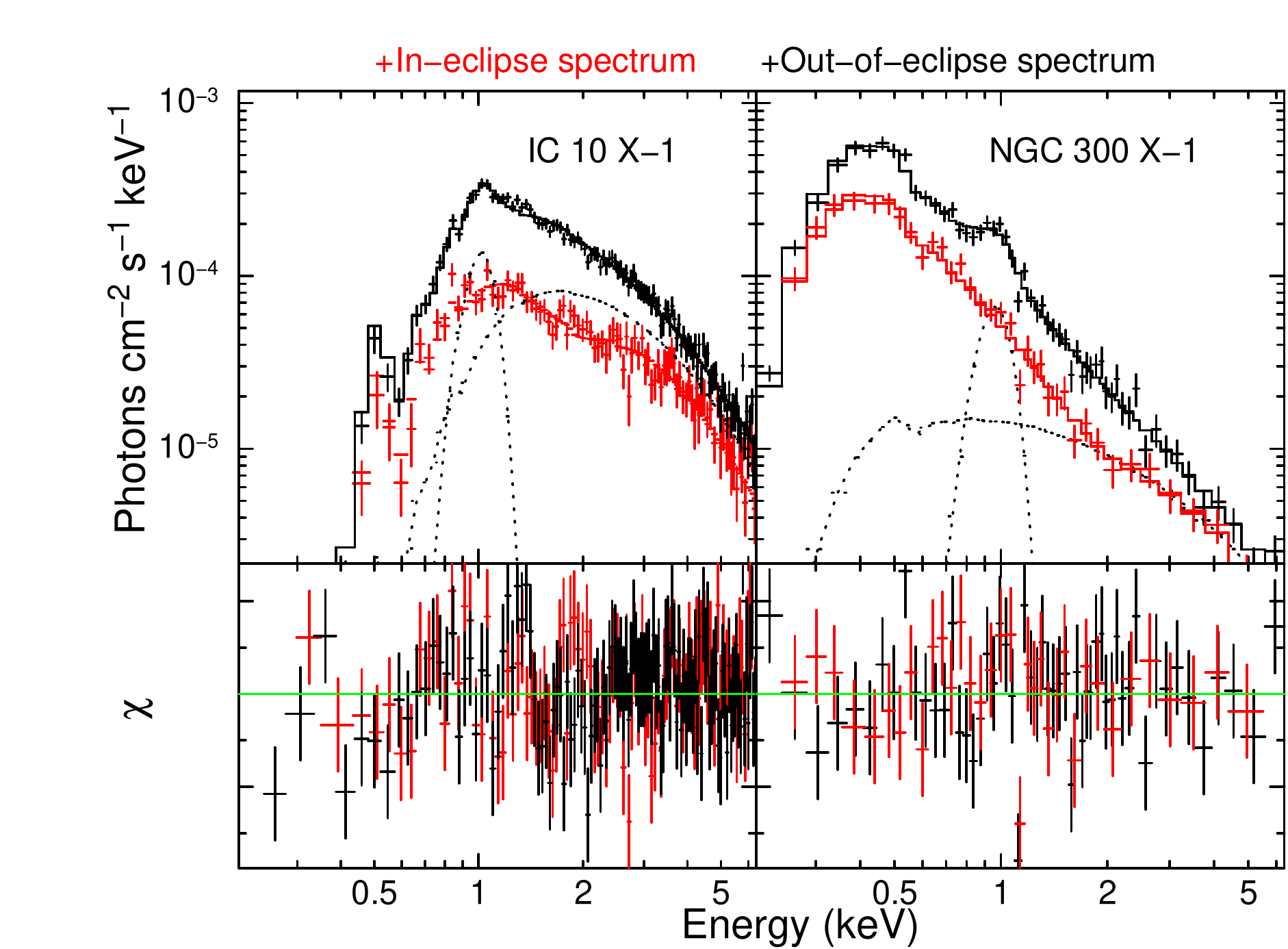}
\caption{The unfolded non-eclipse spectra modelled using Model-1 along with the residual of source IC 10 X-1 and NGC 300 X-1 are plotted in black colour. True eclipse spectra modelled using modified Model-1 \textit{(Tbabs*Tbabs*pcfabs(powerlaw))} are plotted in red color. See the text for more details.}
\label{espec}
\end{figure*}

In IC 10 X-1, the modified Model-1 fitted to both in- and out-of-eclipse spectra show that the out-of-eclipse spectrum is characterized by \textit{diskbb}, \textit{powerlaw} and \textit{gauss} components. However, we find the in-eclipse spectrum contains \textit{powerlaw} component alone, without \textit{diskbb} and \textit{gauss}. The powerlaw index is found to be consistent across these two spectra. The absorption column density ($N_{H}^{pf}$) and covering fraction ($cov\_frac$) during eclipse obtained from \textit{pcfabs} are found to be $7.65^{+0.76}_{-0.69}\times10^{22}$ cm$^{-2}$ and $0.93^{+0.01}_{-0.02}$ respectively. The reduction in flux (0.3$-$10 keV) from $4.14\times10^{-11}$ erg s$^{-1}$ during non-eclipse to $1.1\times10^{-11}$ erg s$^{-1}$ during eclipse is observed. In case of NGC 300 X-1, similar modelling shows that the eclipse spectrum consists solely of the \textit{powerlaw} component, while the \textit{diskbb} and \textit{gauss} components, present in the out-of-eclipse spectrum, no longer observed. We find the $N_{H}^{pf}$ during eclipse to be $9.49^{+3.11}_{-1.91}\times10^{22}$ cm$^{-2}$ and $cov\_frac=0.95^{+0.06}_{-0.10}$ for this source. We find the source flux to be $\sim1\times10^{-12}$ erg s$^{-1}$ during non-eclipse and $\sim5\times10^{-13}$ erg s$^{-1}$ during eclipse. It was not possible to conduct a similar study for M33 X-7 due to the low count rate during eclipse spectrum of Epoch MX14, making it impractical to perform the modelling. In Figure \ref{espec}, we plot the unfolded spectra of non-eclipse period of IC 10 X-1 (left panel) and NGC 300 X-1 (right panel) in black and the corresponding in-eclipse spectra in red. The residuals of respective fits are plotted in the bottom panels. 

Additionally, we attempt to model the eclipse spectra using Model-2, rather than Model-1, to gain further insights into the accretion geometry. However, we found that during the eclipse, neither of the disk models (\textit{diskbb} or \textit{diskpbb}) adequately fit the spectrum. In fact, the eclipse spectrum could only be fitted with an absorbed \textit{powerlaw}, which limits our understanding of the emission observed during the eclipse. We also model the ingress and egress part of the spectrum in IC 10 X-1 and NGC 300 X-1 using Model-1. The IC 10 X-1 spectrum during these period consists of both \textit{diskbb} and \textit{powerlaw} components with absence of only the emission line. The parameters of both these models are found to be consistent with that of the non-eclipse spectrum. The $N_{H}^{pf}$ and $cov\_frac$ during ingress and egress in Epoch IX2 are found to be very similar i.e., $\sim4\times10^{22}$ cm$^{-2}$ and $\sim0.8$ respectively. In case of NGC 300 X-1, similar to true eclipse, we find the ingress and egress spectra to be characterized only the \textit{powerlaw} component. The $N_{H}^{pf}$ and $cov\_frac$ are found to be consistent during ingress and egress with the value of $\sim4\times10^{22}$ cm$^{-2}$ and $\sim0.7$, respectively.

\section{Discussion and Conclusions}
\label{sec5}
We have examined the spectral and timing properties of three extragalactic BH-XRBs namely M33 X-7, NGC 300 X-1 and IC 10 X-1 using observations from \textit{XMM-Newton} and \textit{NuSTAR}. We carry out spectral modelling of non-eclipse observations to gain insights on the accretion disc geometry in these wind-accreting systems. Consequently, we characterize the energy spectra to have a slim accretion disc along with a standard disc. We study eclipse observations by carrying out lightcurve modelling, HR study as well as spectral modelling. Additionally, we estimate the mass range of these BHs by applying continuum-fitting method using relativistic slim-disc model. Based on the results, we deduce the following conclusion.

\subsection{Eclipsing Properties}

In this study, the extragalactic BH-XRBs under investigation exhibit eclipse in their X-ray lightcurves, each lasting for few hours (10.8 hrs in M33 X-7, 12.7 hrs in NGC 300 X-1 and 16.3 hrs in IC 10 X-1). We estimate the size of the companion star (see Section \ref{lcan}) to be $\sim10$ R$_{\odot}$,  $\sim10$ R$_{\odot}$, $\sim18$ R$_{\odot}$ in IC 10 X-1, NGC 300 X-1 and M33 X-7 respectively, using the eclipsing duration. The derived size of WR companions of IC 10 X-1 and NGC 300 X-1 are consistent with the predicted sizes of WR-stars in their pre-supernova phase \citep{2010ApJ...725..940Y}. This confirms that the companion stars are indeed the source of the observed eclipse. The source counts during such eclipse in their \textit{XMM-Newton} lightcurve is found to be energy dependent in NGC 300 X-1. However we could not detect such dependence in IC 10 X-1 and M33 X-7. No-variability in HR during eclipse of M33 X-7 seen in our study is consistent with its \textit{Chandra} lightcurve studied by \cite{2004A&A...413..879P}. Increase in HR seen in NGC 300 X-1 in- and out-of-eclipse is also consistent with the previous studies. 

The spectral study carried out for the eclipsing observations of IC 10 X-1 and NGC 300 X-1 indicates that during eclipse, the thermal disc component gets completely  obscured by the eclipsing matter in both sources (see Section \ref{4.2}). Although we were unable to obtain a suitable spectrum to carry out spectral modelling of M33 X-7, \cite{2006ApJ...646..420P} have carried out spectral modelling using \textit{Chandra} observations, reporting energy spectrum to have consistent parameters in- and out-of-eclipse. The obscuration of disc component with high covering fraction ($\sim0.9$) during eclipse and relatively less covering fraction during entry and exit of the eclipse ($\sim0.75$) in both IC 10 X-1 and NGC 300 X-1 indicate the absorption of soft radiation by additional obscuring material other than the companion star. We speculate such obscuration of soft thermal radiation is caused by stellar wind since both these BHs belong to WR+BH system that has strong stellar winds.

\subsection{Comparison with LMC X-1 and LMC X-3}
\label{sec5.2}
\begin{figure*}
\includegraphics[width=10cm,height=14cm,angle=-90]{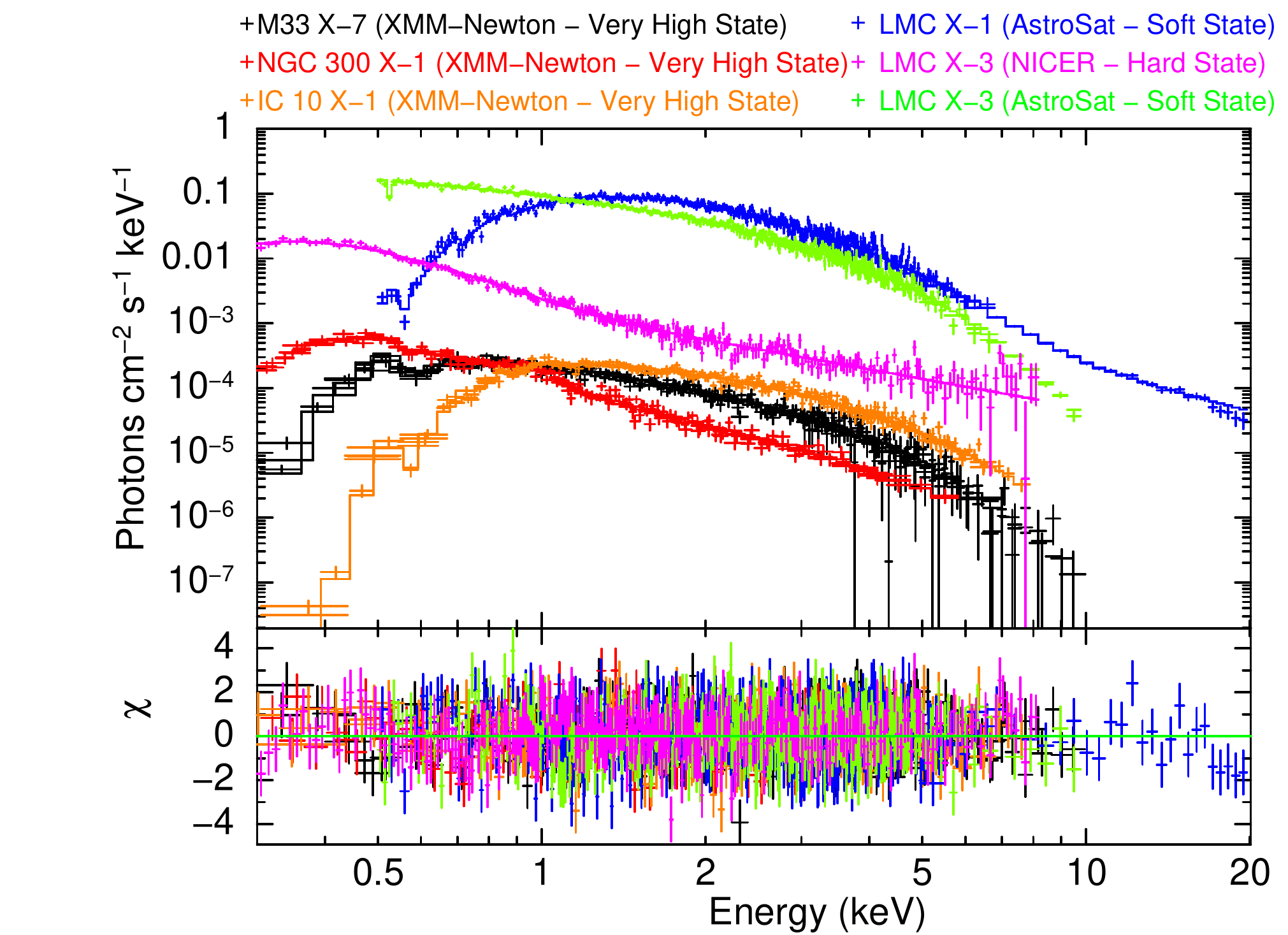}
\caption{Unfolded spectra of all five extragalactic BH-XRBs are plotted in the top panel. Residuals of best-fit of these spectra are plotted in the bottom panel. \textit{EPIC-pn} spectrum and residual of M33 X-7 (black), NGC 300 X-1 (red) and IC 10 X-1 (orange) are modelled with \textit{Tbabs*Tbabs(diskbb+diskpbb)}. \textit{AstroSat-SXT+LAXPC} spectrum of LMC X-1 (blue), hard state spectrum of LMC X-3 from \textit{NICER} (magenta) and soft state spectrum from \textit{AstroSat} (green) are modelled using \textit{Tbabs(ezdiskbb+nthcomp)} with an additional \textit{gauss/apec} whenever required}. 
\label{fig6}
\end{figure*}
The five known extragalactic stellar mass BH-XRBs encompasse LMC X-1 and LMC X-3, in addition to the three sources examined in this study. Previous study of LMC X-1 and LMC X-3 using \textit{AstroSat}, \textit{NICER} and \textit{NuSTAR} observations show that they spend most of the time in thermal dominant soft spectral state \citep{2021MNRAS.501.5457B,2022AdSpR..69..483B}.  LMC X-3 is seen transiting to hard and intermediate states from soft state however, LMC X-1 is found only in soft state. Bolometric luminosity ($0.1-100$ keV) of LMC X-1 is observed to be within $0.07-0.24$ L$_{Edd}$ and that of LMC X-3 within $0.01-0.42$ L$_{Edd}$. In Figure \ref{fig6}, we plot the energy spectra of all five extragalactic BH-XRBs obtained using different instruments. The energy spectra of M33 X-7  (black), NGC 300 X-1 (red) and IC 10 X-1 (orange) represents the source in SPL like spectral state. The energy spectra of LMC X-1 (blue) representing its soft state, and LMC X-3 in both soft (green) and hard state (magenta) are also illustrated in this figure.  

Energy spectra of LMC X-1 and LMC X-3 are found to be characterized by thermal disc component along with Comptonized component, consistent with the standard accretion disc picture. However, as mentioned in Section \ref{sec4.1}, this standard scenario does not describe the energy spectra of M33 X-7, NGC 300 X-1 and IC 10 X-1 in their high luminosity spectral state. Rather, their spectra agree with the previously discussed slim-disc model. It is to be noted that similar to these sources, LMC X-1 also exhibits dominant contribution from \textit{powerlaw} component (or equal contribution from \textit{diskbb} and \textit{powerlaw} component) with a steep energy spectra when its luminosity is close to the Eddington limit \citep{2014MNRAS.445.4259A}. We examine one such observation of LMC X-1 by \textit{XMM-Newton} observed on 21-10-2000. The energy spectrum modelled using \textit{diskbb+powerlaw} yields disc temperature of $\sim0.97$ keV and photon index of $\sim2.8$. The unabsorbed source flux is $1.28\times10^{-9}$ erg cm$^{-2}$ s$^{-1}$ in 0.1$-$100 keV. Considering source distance as 48 kpc and BH mass as 10 M$_{\odot}$, bolometric luminosity during this observations is $3.52\times10^{38}$ erg s$^{-1}$ (i.e., 0.28 L$_{Edd}$). In this high luminosity observation, contribution from \textit{diskbb} is found to be only 12\% with dominant flux coming from non-thermal \textit{powerlaw} component similar to that seen in extragalactic BH-XRBs studied in this work. This demonstrates that the change in spectral state from thermal dominant to SPL-like is associated with increase in luminosity. 

Applying a similar spectral modelling approach used for the three other extragalactic BH-XRBs, we employ \textit{diskpbb} model to fit the SPL-like spectrum of LMC X-1. We observed that spectrum is well described by an absorbed \textit{diskpbb} model without need for additional soft component similar to most of the M33 X-7 spectra. This model resulted in $kT_{in}$ of $1.32\pm0.01$ keV, normalization of $1.29^{+0.07}_{-0.08}$, and p value of 0.5 that hits the lower limit. Striking resemblance of the spectral properties of LMC X-1 and M33 X-7 within a similar luminosity range suggests possibility of identical accretion disc configuration in these sources. Based on these observed spectral properties of all five extragalactic BH-XRBs, we attempt to infer their accretion geometry in different spectral states.

\subsection{Accretion Geometry}
\label{sec5.1}

The preliminary spectral modelling of M33 X-7, NGC 300 X-1 and IC 10 X-1 carried out using \textit{diskbb} and \textit{powerlaw} (see Section \ref{sec4.1}) causes discrepancy due to the extension of \textit{powerlaw} in lower energies. As a result, we opted for alternate modelling scenario, where we tested the presence of optically-thick, advection dominated slim-disc scenario described by \cite{1988ApJ...332..646A} given that, at high mass accretion rate ($\dot{M}>0.1\dot{M}_{Edd}$), the standard \cite{1973A&A....24..337S} model does not hold. This is due to the fact that increase in radiation pressure causes the disc scale height to increase, which results in inner accretion disc to become less radiatively efficient. Consequently, the inner region of accretion disc resembles advection-dominated accretion flow which is optically thick and is referred as `slim-disc'. The use of two thermal components in the modeling is aimed at understanding the accretion scenario in the less-understood steep power-law spectral state, where the \textit{diskbb+powerlaw} model falls short. The choice of the \textit{diskpbb} model along with \textit{diskbb} in place of \textit{powerlaw} is motivated by the similarity of the observed spectral properties of our sources with that of super-Eddington black hole ULXs. ULXs typically exhibit a very steep power-law spectrum and are best described by two thermal components, i.e., \textit{diskbb} and \textit{diskpbb} (see for example, \citealt{2015ApJ...806...65W,2019ApJ...881...38E}). Although such slim-disc is commonly observed in ULXs with super-Eddington luminosities, it is observed only in few sub-Eddington BH-XRBs such as LMC X-1, LMC X-3, XTE J1550-564 and GX 339-4 \citep{2000PASJ...52..133W,2004MNRAS.353..980K,2017ApJ...836...48S}. In these sources, presence of slim-disc is observed only when the source luminosity is high (L$\sim$0.1$-$0.4 L$_{Edd}$), where the energy spectrum resembles that of steep-powerlaw state whose physical origin is unclear. Therefore, we proceed with this approach because the observations of M33 X-7, NGC 300 X-1, and IC 10 X-1 show relatively high luminosities compared to similar persistent Galactic sources.

Results from the spectral modelling show that the spectra of M33 X-7, NGC 300 X-1 and IC 10 X-1 are very well described by a \textit{diskpbb} component along with \textit{diskbb} (see Section \ref{sec4.1}). This suggests the presence of a `hot' slim-disc with temperature within 1$-$2 keV along with a cool standard accretion disc at the outer region having temperature of 0.1$-$0.2 keV (see Table \ref{tab3}). Lower value of $p$ ($<0.66$) in \textit{diskpbb} model, when compared to $0.75$ of standard accretion disc suggests presence of advection in the accretion disc. Such spectral components resemble those observed in ULXs during its ultraluminous spectral state \citep{2009MNRAS.397.1836G,2013MNRAS.435.1758S}. In this state, the high energy thermal component (i.e., \textit{diskpbb}) is attributed to an optically thick, advection dominated accretion disc with an increased scale height. The out-flowing stellar wind from this puffed up disc down scatters some of the high energy radiation which forms the soft spectral component in the energy spectrum resembling a standard multicolor disc black body (i.e., \textit{diskbb}) \citep{2011MNRAS.417..464M}. Although the sources under consideration in this work exhibit similar thermal disc components, their  luminosity remains sub-Eddington (see Table \ref{tab3}). This makes the presence of wind-induced softer component less likely, as wind emission is typically associated with super-Eddington states. Rather, the high normalization value from \textit{diskbb}, approximately 4 times greater than that of \textit{diskpbb} suggests that the soft component corresponds to a thin disc at the outer region of the accretion disc. 

The properties of spectrum and lightcurve observed during eclipse in these sources provide further insights into the accretion geometry. For instance, the non-zero flux observed during true eclipse phase indicates that the emission is coming from object extended farther than the size of companion star. The increase in HR during the lowest flux phase observed in eclipse lightcurve and presence of only the hard emission in eclipse spectrum imply that the extended object must be the `hot' plasma. The complete obscuration of soft disc component during eclipse further supports our understanding of the origin of soft spectral component to be from thin accretion disc at outer accretion region, rather than being associated with the wind. Thus, from these results, we infer that the accretion geometry in these systems possibly involves a `vertically' extended advective flow in the inner region, alongside a standard disc positioned at the outer accretion edge. From the spectral results in Table \ref{tab3}, the radius of the accretion disc can be estimated using the \textit{diskbb} normalization. This calculation gives a radius of 1000$-$6500 km ($<130$ R$_{g}$) for all sources, which is very small compared to the determined companion's radius i.e., $\sim10$ R$_{\odot}$ ($\sim1.4\times10^5$ R$_{g}$) for companion of M33 X-7, $\sim10$ R$_{\odot}$ ($\sim1.3\times10^{5}$ R$_{g}$) for that of NGC 300 X-1 and $\sim18$ R$_{\odot}$ ($\sim2.3\times10^{5}$ R$_{g}$) for IC 10 X-1. As a result, the disc component can easily be obscured by the companion during an eclipse, even if located at the outer edge of the accretion disc. However, not all radiation is blocked during a complete eclipse, as some hard radiation still escapes which is seen in the lightcurve and HR. This further supports that the harder component, likely situated in the inner accretion region based on the temperature and normalization of the \textit{diskpbb} component, has a significant vertical extension.

\begin{figure*}
\includegraphics[width=14cm]{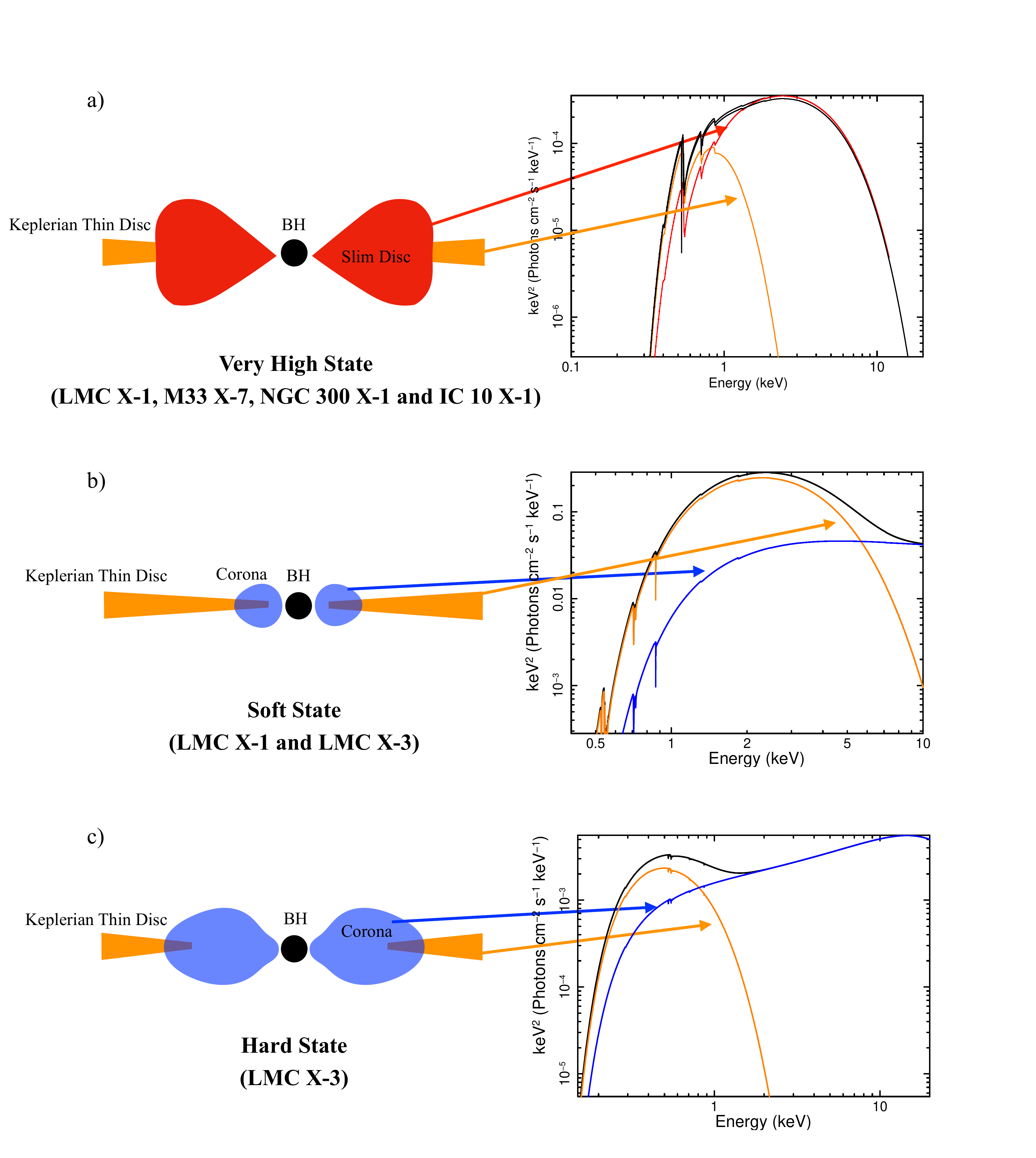}
\caption{The illustration of accretion geometry in extragalactic BH-XRBs during different spectral states. Diagram on the left illustrates the geometry of the accretion disk, while on the right, the observed energy spectrum during respective spectral state is plotted. In Figure (a), very-high/steep powerlaw state is presented, where, along with a cool standard Keplerian disc, an extended slim disc exists in the inner accretion region. The energy spectrum corresponding to this state belongs to M33 X-7, modelled using \textit{diskbb} and \textit{diskpbb}. In Figure (b), the soft spectral state where an accretion disc is present along with a weak corona is shown along with corresponding thermal dominant energy spectrum of LMC X-1 modelled using \textit{diskbb} and \textit{nthcomp}. Figure (c) represents low-hard state where the Comptonization component is dominant in the spectrum. The corresponding spectrum belongs to hard state of LMC X-3 modelled with \textit{diskbb} and \textit{nthcomp}. See the text more details. }
\label{fig8}
\end{figure*}

The observed spectral properties, helped us to infer accretion disc geometry in all five extragalactic BH-XRBs across observed spectral states. M33 X-7, NGC 300 X-1 and IC 10 X-1 shows SPL/Very-High spectral state having two thermal components i.e., \textit{diskbb} and \textit{diskpbb} in their spectra. LMC X-1 exhibits similar spectral state when its luminosity reaches above $0.28$ L$_{Edd}$. While the other three sources are observed in this state through wide range of luminosity i.e., 0.1$-$0.69 L$_{Edd}$. This is illustrated using energy spectrum of M33 X-7 plotted on right in Figure \ref{fig8}(a). On its left, the corresponding accretion geometry consisting of standard thin disc (orange) and a slim disc (red) is shown.  Slim disc is represented as puffed up structure that extends beyond ISCO. Slim discs are optically thick and has aspect ratio, i.e., vertical scale height / radial distance from centre , $H/R \sim1$, hence are thicker than the standard thin disc with $H/R<<1$ \citep{2013LRR....16....1A}. In Figure \ref{fig8}(b), the typical soft state exhibited by LMC X-1 and LMC X-3 is shown by plotting soft state energy spectrum of LMC X-1. LMC X-1 and LMC X-3 are observed to be in soft state for luminosity upto $\sim0.24$ L$_{Edd}$ and $\sim$0.42 L$_{Edd}$ respectively. The accretion geometry is fairly well understood in this state having thin disc extending all the way till $R_{ISCO}$ along with a weak corona (blue). Hard state is shown in Figure \ref{fig8}(c) by plotting energy spectrum of LMC X-3. The accretion geometry consists of thin disc along with dominant corona. LMC X-3 is observed to show hard state when its luminosity is as low as 0.01 L$_{Edd}$. At all luminosities above this value, the source is found either in intermediate or soft spectral state. 

As previously mentioned, the formation of slim disc in super-Eddington sources occurs due to the instability of standard disc as $\dot{M}$ increases. In this scenario, a substantial amount of energy in the inner accretion disc is not cooled radiatively but rather retained as internal energy. This modifies the shape of the spectra from standard multicolor disc blackbody ($kT_{in}\propto r^{-3/4}$) to much broader ($kT_{in}\propto r^{-1/2}$) profile. This prediction however is not applicable for the considered extragalactic BH sources since the deviation from standard accretion picture is observed at luminosity as low as $\sim0.1$ L$_{Edd}$. Such anomalous spectral behaviour is observed in other Galactic BH-XRBs as well. For example, GX 339$-$4 and GRO J1655$-$40 are observed to be deviated from standard accretion scenario at L $\sim0.1$ L$_{Edd}$ \citep{2001ApJ...560L.147K,2017ApJ...836...48S}. The reason for spectral broadening at such low luminosities is unclear. At these luminosities, the accretion disc maybe stabilized by magnetic pressure rather than radiative pressure \citep{2019ApJ...884L..37L}. \cite{2006Natur.441..953M} has found evidence for magnetically powered wind  that stabilizes the disc in Galactic sub-Eddington BH-XRB GRO J1655$-$40. Alternatively, these extragalactic BH-XRBs could be concealed ULXs, much like Galactic BH-XRB Cyg X-3. This source is recently uncovered by \cite{2023arXiv230301174V} to be an ULX having optically thick funnel covering the inner accretion disc, that beams the radiation to be super-Eddington when viewed down the funnel. Therefore, further investigation may be necessary by utilizing X-ray polarimetric data to gain a comprehensive understanding of the accretion disc structure in these extragalactic BH-XRBs.

\subsection{Most massive stellar mass Black hole X-ray Binaries?}

The mass of three extragalactic BHs (M33 X-7, NGC 300 X-1 and IC 10 X-1) are known to reside at the upper limit of the mass spectrum for stellar mass BH-XRBs. As stated in Section \ref{sec1}, the mass of M33 X-7 is well constrained from optical study. The mass of NGC 300 X-1 derived from radial velocity measurement of C IV $\lambda$1550 is subjected to uncertainty associated with the origin of emission line. Despite the striking similarities in the properties of IC 10 X-1 with NGC 300 X-1, the mass of IC 10 X-1 remains unconstrained due to unreliable emission lines  (see Section \ref{sec1}). Therefore, in this work, we attempt to constrain the mass of all three BHs using two methods, first by using its relation with inner disc radius, which is estimated using \textit{diskpbb} normalization. Second, from continuum-fitting method by modelling the energy spectra using relativistic slim disc model \textit{slimbh}. Obtained mass ranges from first method are very broad i.e., mass of M33 X-7, NGC 300 X-1 and IC 10 X-1 to be 7.3$-$51 M$_{\odot}$, 1.3$-$27 M$_{\odot}$ and 3$-$30 M$_{\odot}$ respectively. This is due to large uncertainty associated with inner disc radius value attributed to the uncertainty in disc normalization value, as relativistic effects are not taken into account in \textit{diskpbb} model. Further, we estimate the mass using relativistic slim disc model, which yields much better constrain on mass range i.e., $8.9-14.9$ M$_{\odot}$, 8.7$-$28 M$_{_\odot}$ and 10.2$-$30 M$_{\odot}$ for M33 X-7, NGC 300 X-1 and IC 10 X-1 respectively.  These values are consistent with the dynamically estimated mass. However, these estimated mass values are also subjected to the uncertainties that is associated with distance, since accurate estimate of distance to these systems are not known. Additionally, uncertainties on assumed spin and inclination angle, especially for NGC 300 X-1 and IC 10 X-1 add error to the derived mass range. Furthermore, the mass estimation from continuum-fitting method keeping $a$ and $i$ free would not provide a strong constraint on mass as seen in Figure \ref{fig4}, since the model highly relies on accurate values of these dependent parameters. This indicates the necessity of better constraint on mass using different tools such as from polarization measurement. Nevertheless, the estimated mass range is notable as it is derived for the first time through the implementation of relativistic slim-disc model to the X-ray continuum. 

Mass of these BHs are relatively higher, when compared to LMC X-1 and LMC X-3 which harbours BH of mass $\sim10$ M$_{\odot}$ and $\sim$6 M$_{\odot}$ respectively. Difference in the mass of BH could be due to different mass of progenitor stars in its main sequence phase. The evolutionary scenario involves two massive stars in binary system in which the primary evolves off main sequence to form BH via common envelope phase \citep{1999ApJ...526..152F}. BH in LMC X-1 is understood to be remnant of supernova from progenitor star of mass $\sim60$ M$_{\odot}$ \citep{2017PASP..129i4201H}, meanwhile mass of progenitor star of LMC X-3 is estimated to be 22$-$31 M$_{\odot}$ \citep{2017A&A...597A..12S}. The mass accretion in these two systems are distinct i.e., while LMC X-1 accretes from the stellar wind of the companion, accretion happens via Roche lobe overflow in LMC X-3. This distinction is attributed to the binary evolution of primary and secondary stars in these systems. In LMC X-3, the initial wide orbital separation was reduced during common envelope phase of the binary evolution to $\sim1.7$ days, triggering the mass accretion via Roche-lobe overflow from the companion star mass of $\sim4$ M$_{\odot}$ \citep{2014ApJ...794..154O}. The orbital period of LMC X-1 on the other hand is estimated to be $\sim3.9$ days and the matter is accreted via stellar wind from the companion star of mass $\sim32$ M$_{\odot}$ \citep{2009ApJ...697..573O} which inherited large amount of mass from the primary star of $\sim60$ M$_{\odot}$ during its common envelope phase. Further high mass BH in M33 X-7 ($\sim15$ M$_{\odot}$) originated from the progenitor star of mass 85$-$99 M$_{\odot}$ \citep{2010Natur.468...77V}. Here, the BH accretes mass from the massive companion ($\sim70$ M$_{\odot}$), which is in a 3.45 days orbit via wind-RLOF where, the matter from stellar wind fills the Roche-lobe of donor star \citep{2022A&A...667A..77R}. The presence of more massive BHs in NGC 300 X-1 and IC 10 X-1 suggests that they likely  originated from significantly massive progenitor stars. However, the close orbit of these systems ($\sim32$ hours in NGC 300 X-1 and $\sim35$ hours in IC 10 X-1) presents a challenge to understand their orbital evolution. 

 The estimated mass of these three extragalactic black holes currently stand among the most massive stellar mass black hole X-ray binaries along with few Galactic BH-XRBs. One of the candidate in this list is a transient BH-XRB MAXI J1631$-$479, whose lower limit of mass is estimated to be $15$ M$_{\odot}$ \citep{2023ApJ...944...68R} although not yet confirmed from the dynamical study. Another candidate for massive stellar mass BH-XRB is Cyg X-1, whose mass is recently re-estimated to be $21.2\pm2.2$ M$_{\odot}$ by \cite{2021Sci...371.1046M} and hence currently holds the title of being the most massive stellar mass BH-XRB. Mass of the three extragalactic BH-XRBs could not be constrained to a precise value owing to the uncertain values of its spin, inclination angle and distance. However, if confirmed to be massive, they are crucial for the understanding of the evolution of binary BH systems as they are potential progenitors of binary black hole systems, which is predicted to merge in future that may result in generation of gravitational waves \citep{2014ARep...58..126B}.

\section*{Acknowledgements}
We thank the anonymous reviewers for their useful comments and suggestions that has improved the paper. We acknowledge the financial support of Indian Space Research Organization (ISRO) under AstroSat archival data utilization program Sanction order No. DS$-$2B$-$13013(2)/13/2019$-$Sec.2. This publication uses data from \textit{EPIC} instrument on-board \textit{XMM-Newton} by European Space Agency (ESA). This work also uses data from \textit{NuSTAR} mission by NASA. Also, this research made use of software provided by the High Energy Astrophysics Science Archive Research Center (HEASARC) and NASA’s Astrophysics Data System Bibliographic Services. AN thanks the Group Head, Space Astronomy Group (SAG); Deputy Director, Payload Data Mgmt \& Space Astronomy Area, and Director, U R Rao Satellite Centre (URSC) for encouragement and continuous support to carry out this research.

\section*{Data Availability}

 This paper uses the \textit{XMM-Newton} data available at \url{https://heasarc.gsfc.nasa.gov/docs/xmm/xmmhp_archive.html} and \textit{NuSTAR} at \url{https://heasarc.gsfc.nasa.gov/docs/nustar/nustar_archive.html}.



\bibliographystyle{mnras}
\bibliography{mnras_template} 




\appendix




\bsp	
\label{lastpage}
\end{document}